\documentclass[aps,pre,twocolumn,superscriptaddress,floatfix,longbibliography]{revtex4}
\usepackage{graphicx}
\usepackage{amsmath}

\usepackage{placeins}
\usepackage{bm}
\usepackage{color}
\usepackage{pdfpages}

\begin{document}

\title{Cross-stream migration of active particles}

\author{Jaideep Katuri}
\affiliation{Institut de Bioenginyeria de Catalunya (IBEC), Baldiri I Reixac 10-12, 08028 Barcelona, Spain}
\affiliation{Max-Planck-Institut f\"{u}r Intelligente Systeme, Heisenbergstr. 3, D-70569
Stuttgart, Germany}
\author{William E. Uspal}
\email{uspal@is.mpg.de}
\affiliation{Max-Planck-Institut f\"{u}r Intelligente Systeme, Heisenbergstr. 3, D-70569
Stuttgart, Germany}
\affiliation{IV. Institut f\"ur Theoretische Physik, Universit\"{a}t Stuttgart,
Pfaffenwaldring 57, D-70569 Stuttgart, Germany}
\author{Juliane Simmchen}
\affiliation{Max-Planck-Institut f\"{u}r Intelligente Systeme, Heisenbergstr. 3, D-70569
Stuttgart, Germany}
\author{Albert Miguel-L\'opez}
\affiliation{Institut de Bioenginyeria de Catalunya (IBEC), Baldiri I Reixac 10-12, 08028 Barcelona, Spain}
\author{Samuel S\'anchez}
\email{ssanchez@ibecbarcelona.eu}
\affiliation{Institut de Bioenginyeria de Catalunya (IBEC), Baldiri I Reixac 10-12, 08028 Barcelona, Spain}
\affiliation{Max-Planck-Institut f\"{u}r Intelligente Systeme, Heisenbergstr. 3, D-70569
Stuttgart, Germany}
\affiliation{Instituci{\'o} Catalana de Recerca i Estudis Avancats (ICREA), Pg. Llu{\'i}s Companys 23, 08010, Barcelona, Spain}




\begin{abstract}
For natural microswimmers, the interplay of swimming activity and external flow can promote robust 
\textcolor{black}{directed} 
motion, e.g. propulsion against (‘upstream rheotaxis’) or perpendicular to the direction of flow. These effects are 
generally attributed to their complex body shapes and flagellar beat patterns. Here, \textcolor{black}{using catalytic Janus 
particles as a model experimental system,} we report on a strong directional \textcolor{black}{response} that occurs for 
spherical active particles in a channel flow. \textcolor{black}{The particles align their propulsion axes to be nearly 
perpendicular to both the direction of flow and the normal vector of a nearby bounding surface.} We develop a deterministic 
theoretical model of \textcolor{black}{spherical} \textcolor{black}{micro}swimmers near a planar wall that captures the 
experimental observations. We show how the directional \textcolor{black}{response} emerges from the interplay of shear flow 
and 
near-surface swimming activity.  Finally, adding the effect of thermal noise, we obtain probability distributions for the 
swimmer orientation that \textcolor{black}{semi-quantitatively agree} with the experimental distributions.    
\end{abstract}

\maketitle

\section{Introduction}

In micro-organisms such as bacteria, self-propulsion helps them in the efficient exploration of their surroundings 
to find nutrient rich areas or to swim away from toxic environments. A common feature among natural microswimmers is an 
ability to adjust their self-propulsion in response to local stimuli such as chemical gradients, temperature gradients, 
shear flows or gravitational fields. While the mechanism of some of these responses, such as chemotaxis, is active, 
involving the ability to sense gradients and signals, others are passive, purely resulting from external forces and torques. 
Gravitaxis in Paramecium, for example, is thought to arise solely due to the organism’s fore-rear asymmetry \cite{roberts02}. Recently, there has been substantial effort to develop artificial micro-swimmers which mimic their natural counterparts in many ways \cite{Sanchez2015}. They transduce energy from their local surroundings to engage in self-propulsion and display persistent random-walk trajectories, reminiscent of bacterial run and tumble behaviour \cite{howse07}. They also respond to gravitational fields \cite{Ebbens2013,tenHagen14}, physical obstacles \cite{volpe11,spagnolie12,takagi2014,uspal15a,das15,simmchen16}, and local fuel gradients, seeking fuel-rich regions over depleted ones \cite{Hong2007,baraban13,Peng2015}. A number of applications have been envisioned for these bio-mimetic swimmers, ranging from targeted drug delivery \cite{Patra2013,Gao2014a} to environmental remediation \cite{Soler2014, Gao2014}. In many of these applications, it is inevitable that the artificial microswimmers will encounter flow fields \textcolor{black}{while operating in confined spaces}.  While their swimming behaviour in various complex, static environments has been studied in detail, little is known about their response when they are additionally exposed to flow fields \cite{tao10,zottl12,uspal15b,zottl16}. 

In anticipation of this response, one can look to decades of previous work on biological micro-organisms in flow (\textcolor{black}{see the review papers \cite{pedley92,guasto12} and the references therein.}) The interplay of flow and swimming activity can lead to robust directional response of these microswimmers through a rich variety of mechanisms \cite{marcos12,chengala13,bukatin15,sokolov16}. For instance, in ``gyrotaxis,'' bottom-heavy micro-organisms like algae adopt a stable, steady orientation and swim against the direction of gravity \cite{Kessler85, pedley87, durham09, thorn10, durham11}. Elongated microswimmers such as sperms display ``rheotaxis'', the ability to orient and swim against the direction of flow \cite{bretherton61, hill07, kantsler14, figueroa15, bukatin15, omori16}. However, much of the work with biological microswimmers has focused on swimming in the bulk, and the detailed role of confining boundaries (e.g., in rheotaxis \cite{kantsler14,uspal15b, palacci14, bukatin15, omori16}) is an emerging area of research. 

In this work, using spherical Janus particles as a model system, we demonstrate with experiments and theory 
that spherical particles swimming near a surface can exhibit surprising and counterintuitive spontaneous transverse orientational order.  The particles align almost -- but not exactly -- in the vorticity direction. As we discuss in detail, this slight misalignment is a key experimental observation that allows us to identify and understand the physical mechanism behind the directional alignment. An inactive, bottom-heavy particle in flow 
only rotates as it is carried downstream. However, near-surface swimming activity adds an effective ``rotational friction'' 
to the dynamics of the particle, since it tends to drive the particle orientation into the plane of the wall. The effective ``friction'' damps the particle rotation and stabilises the cross-stream orientation. Finally, the directional alignment, combined with self-propulsion, leads to the cross-stream migration of active particles. The analysis reveals that this mechanism can generically occur for spherical microswimmers in flow near a surface. Our findings exemplify the complex behaviour that can emerge for individual microswimmers from the interplay of confinement and external fields. \textcolor{black}{Moreover, they show that the qualitative character of the emergent behaviour is sensitive to the details of the interactions (e.g., hydrodynamic interactions) between individual microswimmers and bounding surfaces}.

\section{Results}

\subsubsection*{Active particles without external flow}

\begin {figure*}[htp]
\centering
\includegraphics[width = 14.5cm]{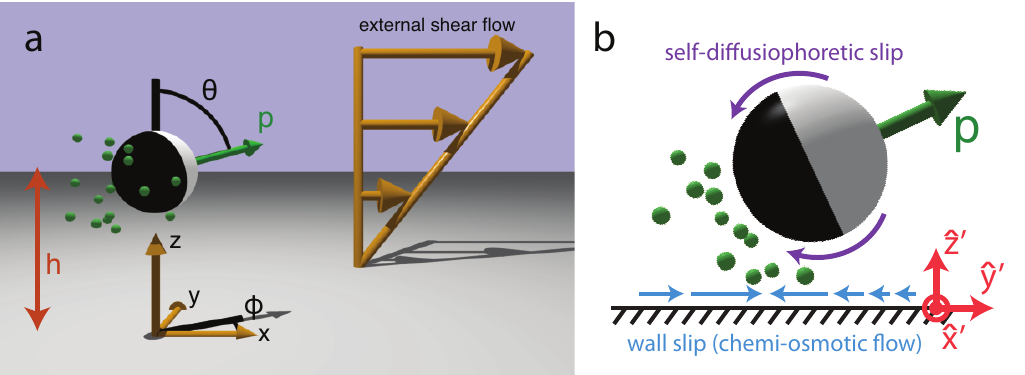}
\caption{\label{fig:schematic} (a) Schematic \textcolor{black}{illustration of the model system}. A particle (white sphere) with axisymmetric coverage by catalyst (black) is driven by an external shear flow (large gold arrows) in the $\hat{x}$ direction near a planar wall (grey). The particle has a height $h$ above the wall and an orientation vector $\mathbf{p}$, which can be specified by the angles $\theta$ and $\phi$. When the particle is active, the cap emits solute molecules (green spheres). (b) Schematic illustrating the self-diffusiophoretic and chemi-osmotic mechanisms that drive the motion of a chemical microswimmer. The self-generated solute gradient (green spheres) drives flows localized to thin boundary layers on the particle surface (magenta arrows) and on the nearby wall (blue arrows). The particle is shown in the ``primed'' frame (red arrows). This frame co-rotates with the particle around the $\hat{z}$ axis, so that the $\hat{z}' = \hat{z}$ and the particle orientation vector $\mathbf{p}$ is always in the plane spanned by $\hat{y}'$ and $\hat{z}'$.}
\end{figure*}

In our experiments, we use silica colloids (1 $\mu m$ radius, or 2.5 $\mu m$ radius where noted) half coated with a thin layer of Pt (10 $nm$) as active particles.  When the particles are suspended in an aqueous $H_2O_2$ solution, the Pt cap catalyses the degradation of $H_2O_2$ while the silica half remains inert. The asymmetric distribution of reaction products creates a concentration gradient along the surface of the particle which induces a phoretic slip velocity, resulting in its propulsion away from the Pt cap. \textcolor{black}{Additionally, particle-generated concentration gradients can induce chemi-osmotic slip on a nearby bounding surface, giving an additional contribution to particle motility. The details of this mechanism, shown schematically in Fig. \ref{fig:schematic}(b),} are comprehensively discussed elsewhere \cite{anderson89,golestanian07}, \textcolor{black}{and are the subject of ongoing research \cite{brown14,ebbens2014,brown2016,brown2017}.}

\begin {figure*}[thp]
\includegraphics[width=16cm]{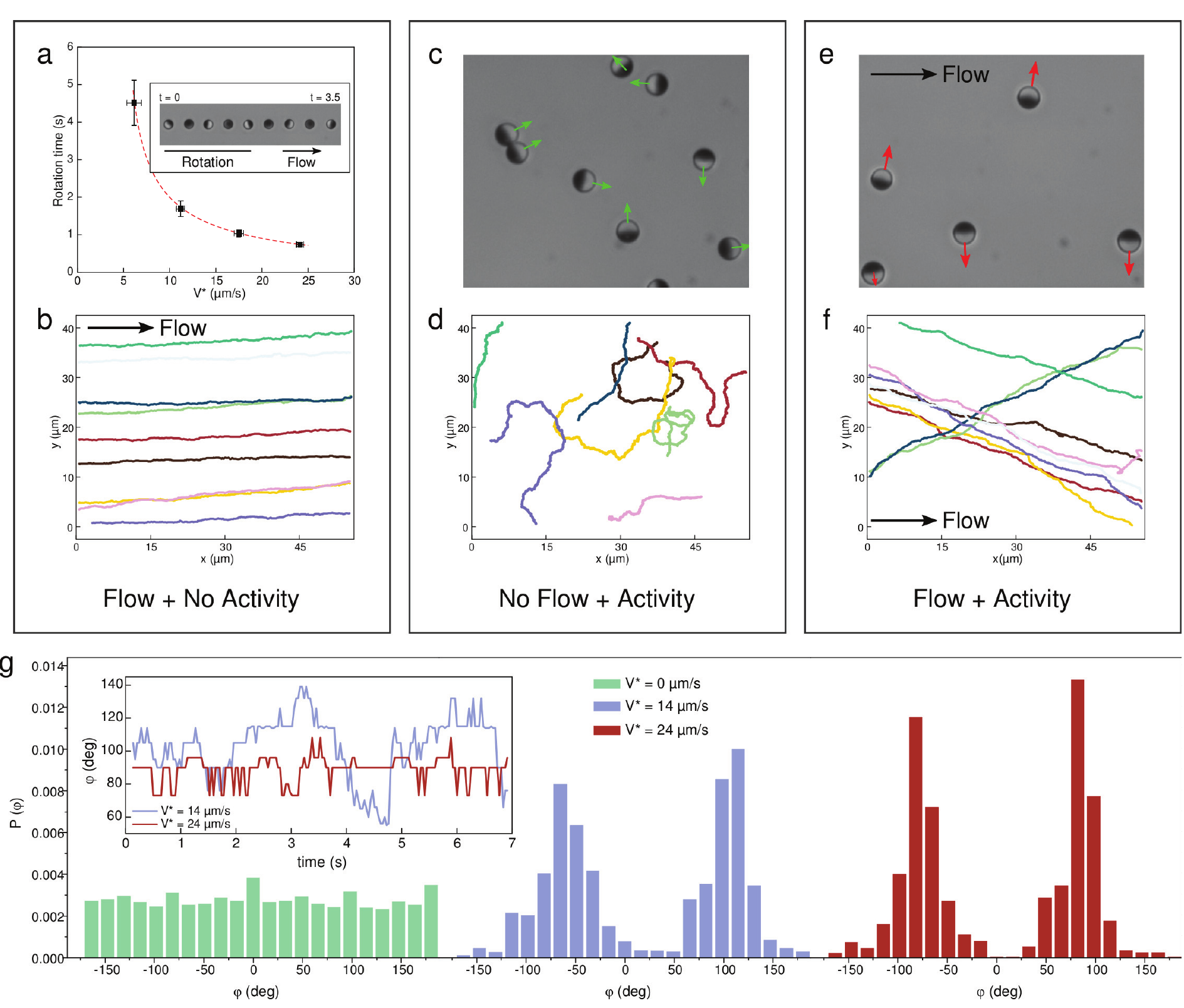}
\caption { \label{fig:imposed_flow_expt} Effect of imposed flow on passive and active colloidal particles.   (a) Passive silica-Pt particles in a flow \textcolor{blue}{($R = 1\;{\mu m}$)}. The plot shows the dependence of the rotation time on flow velocity. $V^*$ is the translational velocity of the passive particle and is used to characterize the flow rate.  \textcolor{black}{The red dashed line is a theoretical scaling derived in  Sect. S4 and fitted to the data.} Inset: Time-lapse images of a passive particle rolling in flow. (b) Tracked trajectories of passive particles in flow ($V^* = 14 \, \mu m/s$).  (c) Optical 
microscopy image capturing the distribution of orientations of active particles \textcolor{black}{($V_{p} \approx 6\;{\mu m}/s$)} without an imposed flow.  (d) Tracked trajectories of active particles without flow. (e) The same system as (c) with an external flow imposed, $V^* = 14 \, \mu m/s$. (f) Tracked trajectories of active particles in flow ($V^* = 14 \, \mu m/s$). (g) The angular probability distributions of active colloidal particles in the absence of an imposed flow, $V^* = 0 \, \mu m/s$ (green), at $V^* = 14 \, \mu m/s$ (blue) and at $V^* = 24  \, \mu m/s$ (red). Inset shows the angular evolution of two different active colloids at imposed flow rates of $V^* = 14 \, \mu m/s$ (blue) and $V^* = 24 \, \mu m/s$ (red).} 
\end{figure*}

When the silica-Pt particles are suspended in water, they quickly sediment 
to the bottom surface as they are density mismatched ($\rho_{SiO_2}$ = 
2.196 $g/cm^3$). They orient with their caps down ($\theta = 0^\circ$, 
\textcolor{black}{where $\theta$ is defined in Fig. \ref{fig:schematic}(a)}) 
due to the bottom heaviness induced by the Pt layer ($\rho_{Pt}$ = 21.45 $g/cm^3$). Addition of $H_2O_2$ introduces activity 
into the system and changes the orientation distribution of the particles. The particles assume an orientation with 
\textcolor{black}{the propulsion axis} parallel to the bottom surface ($\theta = 90^\circ$, see Fig. S1). We have previously 
reported on the dynamics leading to this change in orientation of the active particles \cite{simmchen16}. Briefly, we could show that while the hydrodynamic interactions and the bottom heaviness of the particles tend to drive the particles towards the surface, the wall induced asymmetry of the distribution of the \textcolor{black}{chemical} product and the chemi-osmotic flow along the substrate (see Fig. \ref{fig:schematic}(b)) tend to have the opposite effect, leading to a stable orientation at $\theta \approx 90^{\circ}$. Once parallel to the surface, the particles are confined to a single plane of motion where they propel away from their Pt caps (Fig. \ref{fig:imposed_flow_expt}\textcolor{black}{(c) and (d))} \textcolor{black}{with a typical speed $V_p = 6\;{\mu m/s}$}. Due to the contrast between the dark Pt hemisphere and the transparent silica, we can measure the angular orientations of these particles (see methods and Fig. S2 for details of the tracking process). Within the 2D plane, the particles have no preferred directionality and are diffusive on long time scales.

\subsubsection*{Passive particles in external flow}

Now we seek to characterise the behaviour of these silica-Pt particles in an imposed flow. Initially a suspension of the particles in water is introduced in a square glass capillary (1 $mm$, Vitrocom) connected to a computer controlled microfluidic pump (MFCS-EZ, Fluigent). \textcolor{black}{We allow the particles to sediment to the bottom surface before we impose any external flows.} The desired flow rate in the capillary is maintained by using a flow rate monitor (Flowboard, Fluigent) which is in a feedback loop with the microfluidic pump. We begin by imposing a flow of water (no activity) in the $x$ direction. Close to the non-slipping capillary surface, the flow velocity varies linearly as $v_{flow} = \dot{\gamma} z$ and the particles which are sedimented near the surface experience a shear flow \textcolor{black}{(see Section S3 for a calculation of the flow profile) \cite{bruus08}}.  In terms of their translational behaviour, we observe that the particles act as ``tracers'' and translate in nearly straight lines along the direction of flow \textcolor{black}{(Fig. \ref{fig:imposed_flow_expt}(b))}. The translational velocity of these particles is proportional to the imposed flow rate. We use the translational velocity \textcolor{black}{$V^*$} of these inactive ``tracer''  particles to characterize the flow rate \cite{palacci14}. Before we start the flow, the particles are all in the cap down orientation ($\theta = 0^\circ$). The shear flow induces a torque on the particles and they rotate around \textcolor{black}{the axis of flow vorticity $\hat{y}$} as they translate in flow (See SM Video 1). The rotation speed of the particles is also dependent on the flow rate \textcolor{black}{(Fig. \ref{fig:imposed_flow_expt}(a))}, with higher flow rates leading to faster rotation. \textcolor{black}{Via a simple model  for particle rolling developed in Section S4 of the SM, we predict that the rotational period $\tau = 2 \pi/\sqrt{a V^{*,2} + b}$, where $a$ and $b$ are fitting parameters. This relation shows excellent fit to the data (Fig. 2a, red curve), and from the fitted $a$, we extract $\dot{\gamma} \approx 1.22 \, V^{*}/R$.}


\subsubsection*{Active particles in external flow}

We then start a flow of $H_2O_2$ to introduce activity into the system. We notice that the dynamics of particle behaviour in flow is drastically influenced in the presence of activity. Firstly, the particles stop rolling and reach a stable orientation parallel to the bottom surface ($\theta = 90^{\circ}$). This is similar to previously observed behaviour for the particles in the absence of flow \cite{simmchen16}. More surprisingly, the particles also evolve to a stable orientation  that is nearly perpendicular to the direction of imposed flow ($\phi \approx 90^{\circ}$ or $-90^{\circ}$) (Fig. \ref{fig:imposed_flow_expt}(e)). While the particles continue to translate in $x$ due to the imposed flow, they also have the self-propulsion velocity $V_p$ away from their Pt caps. A combination of these effects results in the cross-streamline migration of self-propelled particles, i.e,  \textcolor{black}{migration of particles in the y-direction, perpendicular to the flow along the x-direction.} 
Typical trajectories of cross-stream migrating particles are presented in Fig. \ref{fig:imposed_flow_expt}(f). Further, we observe that the stability of cross-stream migration (due to the stability of the steady orientation angle \textcolor{black}{$\phi^{*}$} perpendicular to the direction of flow) is dependent on the flow rate, with higher flow rates resulting in a stronger alignment effect.  The inset of Fig. \ref{fig:imposed_flow_expt}(g) shows the angular evolution of two self-propelled particles in imposed flows of $V^* = 14 \, \mu m/s$ and $V^* = 24 \, \mu m/s$. The deviations away from the $\phi \approx  90^\circ$ positions occur more frequently and \textcolor{black}{at larger amplitude} than for particles in lower flow rates. 

In order to study this effect at a population scale we flow a suspension of self-propelled particles with $H_2O_2$ and record the angular orientations and positions of every particle in each frame, which allows us to determine the probability distribution of $\phi$ in the system. These are plotted in Fig. \ref{fig:imposed_flow_expt}(g) for two different flow rates and compared to the system of self-propelled particles without any imposed flow. In the absence of flow, the distribution is nearly flat, indicating the lack of preference for any orientation $\phi$, and at long time scales, the particle behaviour is purely diffusive. However, in the case of imposed flow, we observe distinct peaks that appear at $\phi \approx  90^\circ$ and $-90^\circ$. These correspond to the particles exhibiting the cross-stream behaviour. These peaks also become sharper when we increase the flow rate, as can be seen in the distributions for $V^* = 14 \, \mu m/s$ and $V^* = 24 \, \mu m/s$. \textcolor{black}{In both cases,} closer observation of the angular probability distributions reveals a small bias of orientations in the direction of flow for particles migrating \textcolor{black}{across flow streamlines}. 
Since the particles are subject to Brownian fluctuations, they can occasionally also orient with or against the flow ($\phi \approx 0^\circ$ or $180^\circ$). This leads to an intermittent state where the particles ``tumble'' in the direction of flow before recovering their $\phi \approx  \pm 90^\circ$ orientation and the cross-streamline behaviour. The recovered orientation of these particles can be different from their initial states, changing the direction of particle migration (e.g., from $+y$ to $-y$) (see \textcolor{black}{Fig. S5}).   

\textcolor{black}{Apart from the imposed flow rate $V^{*}$, we find that the stability of the $\phi^{*}$ is also dependent on the particle radius. Using particles of larger radius ($2.5\;{\mu m}$), we show that the distribution of $\phi$ is narrower around $\phi = \pm 90^{\circ}$ as compared to the distribution for $R = 1\;{\mu m}$ particles for identical $V_{p}$ and $V^{*}$ (Fig. S6). Since the effect of Brownian noise is significantly lower on the larger particles, we also observe that they seldom switch their migration direction within the width of our capillary (Fig. S7).}  

\begin{figure*}[htp]
\centering
\includegraphics[width=13.5cm]{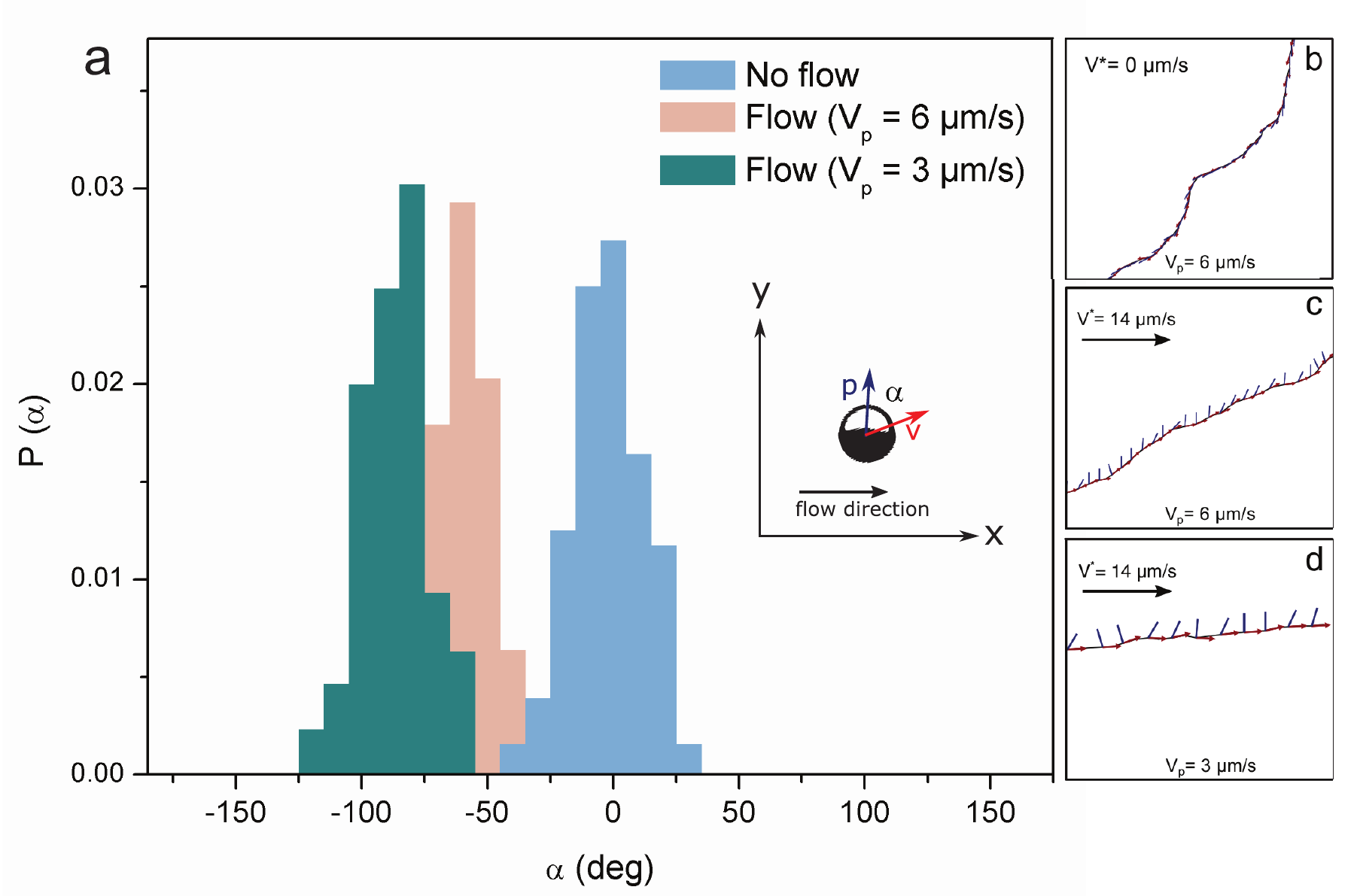}
\caption {\label{fig:alphas} The dependence of cross-stream migration on $V_p$. (a) The probability distributions of $\alpha$ for a particle in the absence of flow with $V_p = 6 \, \mu m/s$\textcolor{black}{, as well as} with an imposed flow corresponding to $V^* = 14 \, \mu m/s$ for particles of $V_p = 6 \, \mu m/s$ and $V_p = 3 \, \mu m/s$. \textcolor{black}{(b, c, and d) Tracked orientation vectors (line segments) and instantaneous velocities (arrows) for trajectories at different flow and self-propulsion speeds.}}   
\end{figure*}  

We can further control the behaviour of particle migration by tuning the self-propulsion velocity $V_p$ of the particles. \textcolor{black}{Firstly, we find that higher propulsion velocities dampen the fluctuations around $\phi^{*}$ due to higher activity (See Fig. S8). Secondly, the propulsion velocity also controls the ``slope'' of the cross-stream migration. In order to quantify this, we} define $\alpha$ to be the offset between the orientation vector $\mathbf{p}$ and the \textcolor{black}{tracked} velocity vector $\mathbf{v}$ of the particle. For a self-propelled particle in the absence of flow, the particles translate in the direction of the orientation vector $\mathbf{p}$ as the particles propel away from their Pt caps (\textcolor{black}{i.e.,} $\alpha = 0^\circ$), \textcolor{black}{as shown in} Fig. \ref{fig:alphas}(a and \textcolor{black}{b}). However, with an imposed flow, the translational direction differs from the orientation vector\textcolor{black}{,} and the offset $\alpha$, for a given flow rate, is determined by the $V_p$ of the particles. For particles with $V_p = 6 \, \mu m/s$ the probability distribution function of $\alpha$ has a peak around $-64^\circ$, whereas for a particle with $V_p = 3 \, \mu m/s$, $\alpha$ is peaked at $-83 ^\circ$ (Fig. \ref{fig:alphas}(a, c and d)).
\textcolor{black}{The offset $\alpha$ can be rationalized as a contribution of the transverse propulsion induced by the cross-stream orientation and the longitudinal advection by flow. The $\alpha$ should then simply be given by $\alpha = \arctan (V^{*}/V_{p})$. Substituting the experimental values for $V_{p}$ and $V^{*}$, we get $\alpha = 66.9^{\circ}$ and $\alpha = 78.3^{\circ}$, which agree reasonably well with the observed values.  This clear separation in peaks of $\alpha$ for different propulsion velocities could eventually be used for the separation of particles based on activity. }

\subsubsection*{Construction of theoretical model}

\textcolor{black}{The experimental observations can be qualitatively captured and understood within a generic mathematical model of swimming near a surface in external flow \cite{uspal15b}. In this model, we construct dynamical equations for the height and orientation of a heavy spherical microswimmer by exploiting physical symmetries and the mathematical linearity of Stokes flow. As detailed below, mathematical analysis of these equations reveals that qualitatively distinct steady state behaviors, including cross-stream migration, can emerge from the interplay of external shear flow, near-surface swimming, and gravity. We note that the analysis does \textit{not} depend on a particular mechanism of self-propulsion (e.g., self-diffusiophoresis, self-electrophoresis, or mechanical propulsion by motion of surface cilia); it is, in that sense, \textit{generic}. Accordingly, our major theoretical findings are, first, that there is a physical mechanism that can produce the surprising transverse orientational order observed in experiments, and, secondly, that this mechanism can \textit{generically} occur for a spherical, \textcolor{black}{axisymmetric} microswimmer exposed to flow near a bounding surface.}

\textcolor{black}{We now proceed to construction and analysis of the dynamical equations. As discussed in the Methods section, the flow in the suspending fluid is characterised by a low Reynolds number, and hence governed by the Stokes equations. Since these equations are linear,} the contributions of external \textit{\textbf{f}}low, \textit{\textbf{g}}ravity, and \textit{\textbf{s}}wimming \textcolor{black}{to the particle translational and angular velocities} can be calculated independently and superposed: $\mathbf{U} = \mathbf{U}^{(f)} + \mathbf{U}^{(s)} + \mathbf{U}^{(g)}$ and $\mathbf{\Omega} = \mathbf{\Omega}^{(f)} + \mathbf{\Omega}^{(s)} + \mathbf{\Omega}^{(g)}$. \textcolor{black}{The velocity of the orientation vector is determined by  $\dot{\mathbf{p}} = \mathbf{\Omega} \times \mathbf{p}$. We will generally describe $\mathbf{p}$ in Cartesian coordinates, but will sometimes find it useful to use the spherical coordinates $\mathbf{p} = (\sin(\theta) \cos(\phi), \sin(\theta) \sin(\phi), \cos(\theta))$. Note that $|\mathbf{p}| = 1$.} 

\begin{figure*}[thp]
\centering
\includegraphics[width = 0.9 \textwidth]{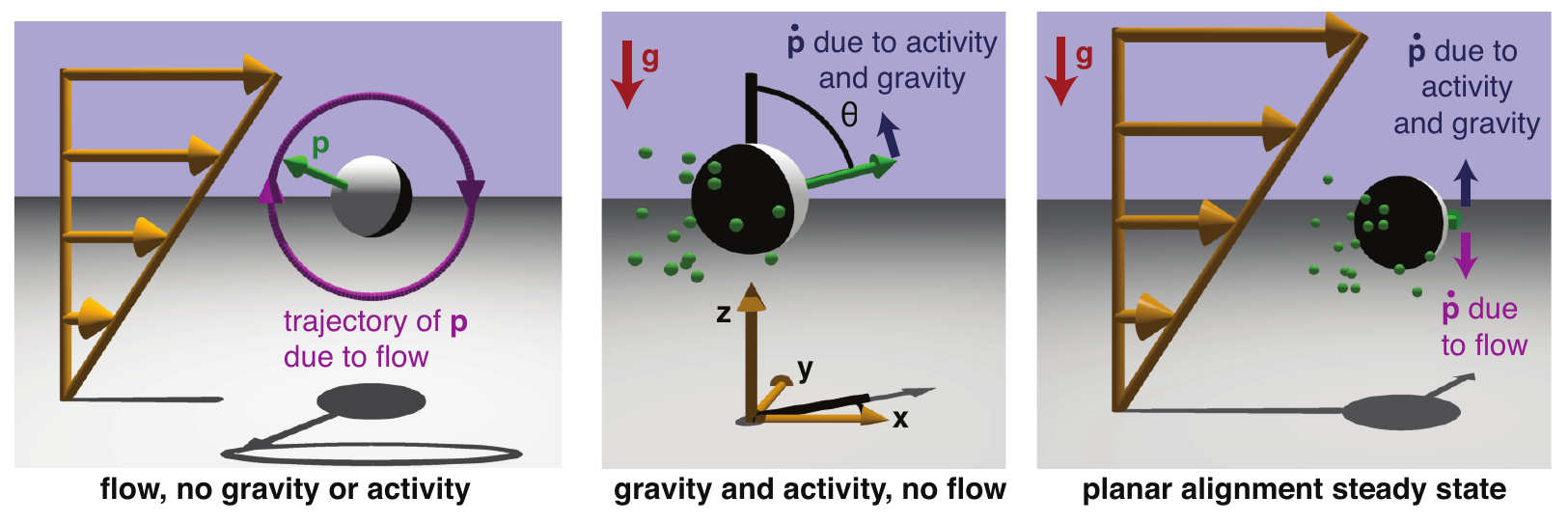}
\caption{\label{fig:decomposition} \textcolor{black}{(left) A neutrally buoyant, inactive sphere driven by an external shear flow (gold arrows) near a planar wall. Since the external flow spins the particle around the vorticity axis (i.e., the normal to the shear plane), the tip of the particle orientation vector $\mathbf{p}$ traces a circular path (magenta). (middle) An active, heavy sphere in quiescent fluid. Since the wall is uniform, and the geometry and activity profile of the particle are axisymmetric, the tip of the orientation vector can only rotate in the $\hat{\theta}$ direction, i.e., directly towards or away from the wall. (right) The \textit{planar alignment} steady state, where $p_{z}^{*} = 0$, but the particle orientation vector has non-zero components $p_{x}^{*}$ and $p_{y}^{*}$ in the flow and vorticity directions, respectively. For planar alignment, the component $p_{x}^{*}$ in the flow direction can be either upstream ($p_{x}^{*} > 0$) or downstream ($p_{x}^{*} < 0$), as determined by the function $\Omega_{x'}^{\textcolor{black}{(ax)}}(p_{z}, h/R)$; the downstream case is shown. All contributions to $\dot{\mathbf{p}}$ are in the $\hat{z}$ direction (see Eqs. \ref{dot_px}-\ref{dot_h} and the discussion in the main text). At a certain angle $\phi^{*}$, all contributions to $\dot{\mathbf{p}}$ balance, as shown by the arrows, so that $\dot{\mathbf{p}} = 0$. Note that this fixed point always occurs in pairs related by mirror symmetry across the shear plane; we show the state with $p_{y} > 0$.}}
\end{figure*}

\textcolor{black}{In order to calculate the contribution of the external flow to the particle velocity, we consider a neutrally buoyant and inactive sphere of radius $R$ in shear flow at a height $h$ above a uniform planar wall.  Shear spins the particle around the vorticity axis $\hat{y}$, as shown in the left panel of Fig. \ref{fig:decomposition}. This is because the flow is faster near the upper surface of the particle (i.e., the surface farther away from the wall) than near the bottom surface.  Accordingly, the particle has angular velocity $\mathbf{\Omega}^{(f)} = \frac{1}{2} \dot{\gamma} f(h/R) \hat{y}$. Additionally, the particle is carried downstream with velocity $\mathbf{U}^{(f)} =  \dot{\gamma} h g(h/R) \hat{x}$.  The functions $f(h/R)$ and $g(h/R)$ represent the influence of hydrodynamic friction from the wall \cite{goldman67}; they are provided in \textcolor{black}{SM Sect. S8}.  Due to the spinning motion of the particle, the tip of the vector $\mathbf{p}$ traces a circle in the shear plane:}
\begin{equation}
\label{eq:pf}
\dot{\mathbf{p}}^{(f)} \equiv \mathbf{\Omega}^{(f)} \times \mathbf{p} =  \frac{1}{2} \dot{\gamma} f(h/R) \left[ p_{z} \hat{x}  - p_{x} \hat{z} \right] 
\end{equation}
The component $p_{y}$ is a constant determined by the initial orientation of the particle. The radius of the \textcolor{black}{circular orbit} is $\sqrt{1 - p_{y}^{2}}$, and the speed of the tip \textcolor{black}{$|\dot{\mathbf{p}}^{(f)}|$} is proportional to the circle radius.

\textcolor{black}{Now we consider a heavy, active particle moving in quiescent fluid (no external flow) near the same surface. The particle has an \textbf{\textit{ax}}isymmetric geometry and surface activity profile; accordingly, we define  $\mathbf{U}^{(ax)} \equiv \mathbf{U}^{(s)} + \mathbf{U}^{(g)}$ and $\mathbf{\Omega}^{(ax)} \equiv \mathbf{\Omega}^{(s)}  + \mathbf{\Omega}^{(g)}$. This system, depicted in the middle panel of Fig. \ref{fig:decomposition}, has a plane of mirror symmetry defined by $\hat{z}$ and $\mathbf{p}$. Accordingly, translation of the particle is restricted to this plane, and $\mathbf{p}$ is restricted to remain within this plane (i.e., $\dot{\phi} = 0$), although $\mathbf{p}$ can rotate towards or away from the wall ($\dot{\theta} \neq 0$). We write $\dot{\mathbf{p}}^{(ax)} = - \Omega_{x'}^{(ax)}(\theta, h/R) \, \hat{\theta}$, where we have defined a new, ``primed'' frame that has $\hat{y}'$ and $\hat{z}'$ in the plane of mirror symmetry, with $\hat{z}' = \hat{z}$. This frame is convenient for calculations because $\mathbf{\Omega}^{(ax)}$ is strictly in $\hat{x}'$  (see Fig. \ref{fig:schematic}(b)). The function $ \Omega_{x'}^{(ax)}(\theta, h/R)$ incorporates the effect of bottom-heaviness, as well as interactions with the wall (e.g., hydrodynamic interactions) that originate in swimming activity. Both $ \Omega_{x'}^{(ax)}(\theta, h/R)$  and $\mathbf{U}^{(ax)}(\theta, h/R)$ depend on the particle design and the model for the propulsion mechanism. In order to show that our analysis is generic in the sense discussed above, we will leave them unspecified until later in the text. Since the particle is axisymmetric and the \textcolor{black}{wall} is uniform, these functions have no $\phi$ dependence.}  

\textcolor{black}{Now we superpose all contributions and obtain equations for $\dot{\mathbf{p}} = \mathbf{\Omega} \times \mathbf{p}$ and $\dot{h} = U_{z}$ in the following form}:
\begin{eqnarray}
\label{eq:dynamical_eqns}
\dot{p}_{x} &=&  \, \frac{1}{2} \dot{\gamma} p_{z} 
f(h/R) -\frac{\Omega_{x'}^{\textcolor{black}{(ax)}}(p_{z}, 
h/R) p_{x} p_{z}}{\sqrt{1 - p_{z}^{2}}}\,  \,,
\label{dot_px} \\
\dot{p}_{y} &=& \,- \frac{\Omega_{x'}^{\textcolor{black}{(ax)}}(p_{z}, h/R) p_{y} p_{z}}{\sqrt{1 - p_{z}^{2}}} \label{dot_py} \\
\dot{p}_{z} &=&   -\frac{1}{2} \dot{\gamma} p_{x} 
f(h/R) + \Omega_{x'}^{\textcolor{black}{(ax)}}(p_{z}, h/R) \sqrt{1 - p_{z}^2} \,, \label{dot_pz} \\
\dot{h}_{~} &=& U_z = \, U_{z'}^{\textcolor{black}{(ax)}}(p_{z}, h/R)\,\label{dot_h}.
\end{eqnarray}
\noindent \textcolor{black}{Note that the axisymmetric contributions are evaluated in the ``primed'' frame, which is defined to co-rotate with the particle, and we used $\hat{\theta} = \cos(\theta) \cos (\phi) \hat{x} + \cos(\theta) \sin(\phi) \hat{y} - \sin(\theta) \hat{z}$.} The components of the translational velocity $\mathbf{U}$ in the $x$ and $y$ directions are determined by $\mathbf{p}$ and $h$ as:
\begin{eqnarray}
 U_{x} &=& \dot{\gamma} \textcolor{black}{h} g(h/R) + \frac{U_{y'}^{\textcolor{black}{(ax)}}(p_{z}, h/R) p_{x}} {\sqrt{1 - p_{z}^{2}}} \\
\label{uy}
 U_{y} &=& \frac{U_{y'}^{\textcolor{black}{(ax)}}(p_{z}, h/R) p_{y}} {\sqrt{1 - p_{z}^{2}}}.
\end{eqnarray}

\subsubsection*{Steady state solutions}

\textcolor{black}{Eqs. \ref{dot_px}-\ref{uy} fully describe a deterministic active sphere in external shear flow near a planar surface.} Now we look for the fixed points $(\mathbf{p}^{*}, h^{*})$ of Eqs. \ref{dot_px}-\ref{dot_h}. A fixed point is a particle configuration in which the particle translates along the wall with a steady height and orientation, i.e., $\dot{\mathbf{p}} = 0$ and $\dot{h} = 0$ \cite{uspal15b}. We find that the system has three fixed points, shown schematically in \textcolor{black}{Fig. S10}. Of these three, \textit{planar alignment} shows excellent qualitative agreement with the experiment observations: the particle orientation is within the plane of the wall ($p_{z}^{*} = 0$), and has non-zero components in \textit{both} the flow and vorticity directions. These criteria are not satisfied by the other two fixed points. \textcolor{black}{Slight misalignment from the vorticity axis is therefore a key experimental observation that discriminates between the steady states predicted by theory.}  For planar alignment, the \textcolor{black}{streamwise} component \textcolor{black}{$p_{x}^{*}$ is 
\begin{equation}
\label{eq:px_star}
p_{x}^{*} = \cos(\phi^{*}) = \frac{2 \Omega_{x'}^{(ax)}(p_{z}^{*} = 0, h^{*}/R)}{\dot{\gamma} f(h^{*}/R)},
\end{equation}
which can be} either downstream ($p_{x}^{*} > 0$), \textcolor{black}{in agreement with our experimental observations,} or upstream ($p_{x}^{*} < 0$), depending on the sign of $\Omega_{x'}^{\textcolor{black}{(ax)}}(p_{z}^{*} = 0, h^{*})$. Due to the mirror symmetry of Eqs. \ref{dot_px}-\ref{dot_h} with respect to $p_{y} = 0$, this fixed point always occurs in pairs $(p_{x}, \pm p_{y}^{*}, 0, h^{*})$. Further mathematical details are provided in the SM.


\textcolor{black}{In order to understand why planar alignment is a steady state, we consider the conditions for the contributions of shear flow, activity, and gravity to the angular velocity to cancel out at some $\mathbf{p}^{*}$. We recall that the axisymmetric contributions to $\dot{\mathbf{p}}$ are in the $\hat{\theta}$ direction. For most orientations $\theta$ and $\phi$, $\hat{\theta}$ will have a component in the vorticity direction $\hat{y}$. However, from Eq. \ref{eq:pf}, we see that the shear contribution \textit{never} has a $\hat{y}$ component. As a way out of this dilemma, we see from the definition of $\hat{\theta}$ that when $\theta = 90^{\circ}$  (i.e., $p_{z} = 0$), the $\hat{x}$ and $\hat{y}$ components of $\hat{\theta}$ both vanish, and $\dot{\mathbf{p}}^{(ax)}$ is strictly in  the $\hat{z}$ direction. Likewise, when $\theta = 90^{\circ}$, $\dot{\mathbf{p}}^{(f)}$ is strictly in $\hat{z}$. This suggests the possibility that all contributions can cancel out for $\theta^{*} = 90^{\circ}$ and some unknown $\phi^{*}$.  \textcolor{black}{Now we consider the role of $\phi$}. The axisymmetric contributions \textcolor{black}{$\dot{\mathbf{p}}^{(ax)}$} have no dependence on $\phi$, as discussed previously. The contribution from shear does depend on $\phi$, \textcolor{black}{as can be seen by substituting $p_{z} = 0$ and $p_{x} = \cos(\phi)$ into Eq. \ref{eq:pf}}.  Therefore, the sign and magnitude of the shear contribution can be ``tuned'' via the angle $\phi$ to cancel out the axisymmetric contributions \textcolor{black}{to $\dot{\mathbf{p}}$}  (provided these are not too large).}   \textcolor{black}{Finally, we consider the height of the particle. Since shear does not contribute to vertical motion, the particle must obtain a steady height ($\dot{h} = 0$) through the combined effects of near-surface swimming activity and gravity.}


\textcolor{black}{As an additional note, if we assume that the steady height of the particle is not significantly affected by flow rate, Eq. \ref{eq:px_star} predicts that the steady state orientation approaches the vorticity axis as the flow rate is increased, i.e., $\cos(\phi^{*}) \sim 1/V^{*}$.  Accordingly, we perform further experiments with $5\;{\mu m}$ diameter particles, which allow for excellent resolution of their orientation (see Fig. S2). We performed the experiments at six different flow rates in the range accessible with the current experimental setup, $V^* = 5\;{\mu m/s}$ to $V^* = 75\;{\mu m/s}$, for particles with $V_p = 6\;{\mu m/s}$. In Fig. S11, we plot $\cos|\phi^{*}|$ against $V^*$ and find that the data recovers the predicted asymptotic behavior.}

\subsubsection*{Linear stability analysis}

\textcolor{black}{It is not enough to find a fixed point that corresponds to experimental observations; we must also consider its stability.} Since, experimentally, the active particles \textcolor{black}{spontaneously adopt a steady cross-stream orientation}, the fixed point should be a stable attractor for active particles. Secondly, since the inactive particles are  observed to  \textcolor{black}{continuously rotate} in the experiments, the fixed point should be associated with closed orbits for inactive particles. \textcolor{black}{In the following, we carry out a linear stability analysis and show that our model meets both criteria, provided a certain condition on the particle/wall interaction is satisfied.} 


 We define the generalized configuration vector $\mathbf{f} \equiv (\mathbf{p}, h/R)$, and $\tilde{\mathbf{f}}$ as a small perturbation away from the fixed point: $\mathbf{f} = \mathbf{f}^{*} + \tilde{\mathbf{f}}$.  We obtain the linearized governing equations $\dot{\tilde{{\mathbf{f}}}} = \mathbf{J} \tilde{\mathbf{f}}$, where the Jacobian matrix $J_{ij} = \frac{\partial \dot{f}_{i}}{\partial f_{j}} \biggr\rvert_{\textcolor{black}{\mathbf{f}^{*}}}$ is given in detail in the SM.  As a simplifying approximation, we take the particle to have a constant height $h = H$, so that $\dot{\tilde{h}} = 0$. This approximation is motivated by the experimental observation that the particles never leave the microscope focal plane. Having made this approximation, we can find an intuitive analogy between our system and a damped harmonic oscillator (see SM):

\begin{figure*}[bhp]
\centering
\includegraphics[width=15.cm]{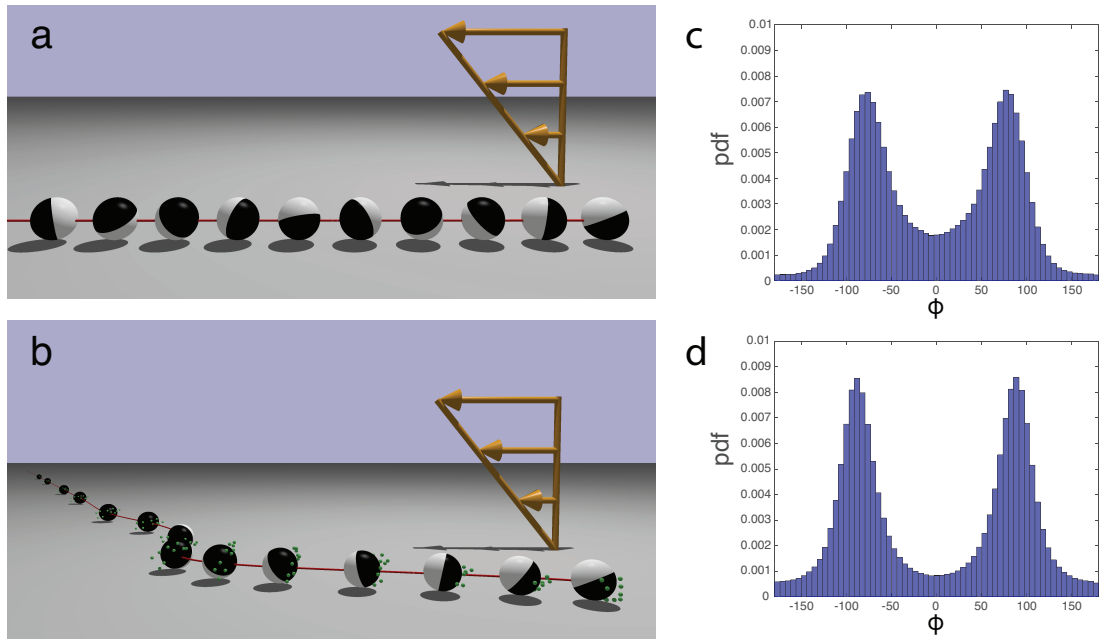}
\caption{\label{fig:traj_pdfs_combined} (a)  Numerically computed trajectory of a passive bottom-heavy particle near a wall (grey) and driven by an external shear flow (gold arrows) with dimensionless strength $\dot{\gamma} R/U_{0} = 0.1$ (concerning dimensionless parameters, see Methods). The initial condition of the particle is $\theta_{0} = 30^{\circ}$ and $\phi_{0} = 315^{\circ}$, and the particle height is fixed as $h/R = 1.2$. The particle rotates and is carried downstream by the external flow with no cross-streamline migration. (b) Numerically computed trajectory of an active Janus particle with the external flow strength, particle materials, and initial conditions as in (a). The particle rotates so that its inert face points largely in the $-\hat{y}$ direction, with a slight downstream orientation ($p_{x} > 0$). With this steady orientation, the particle swims across flow streamlines as it moves downstream.  (c) Probability distribution function for $\phi$, for a catalytic Janus particle in a shear flow with $\dot{\gamma}R/U_{0} = 0.1$. (d) Probability distribution function for the same particle with $\dot{\gamma}R/U_{0} = 0.5$.}
\end{figure*}

 \begin{figure*}[thp]
 \centering
\includegraphics[width=17cm]{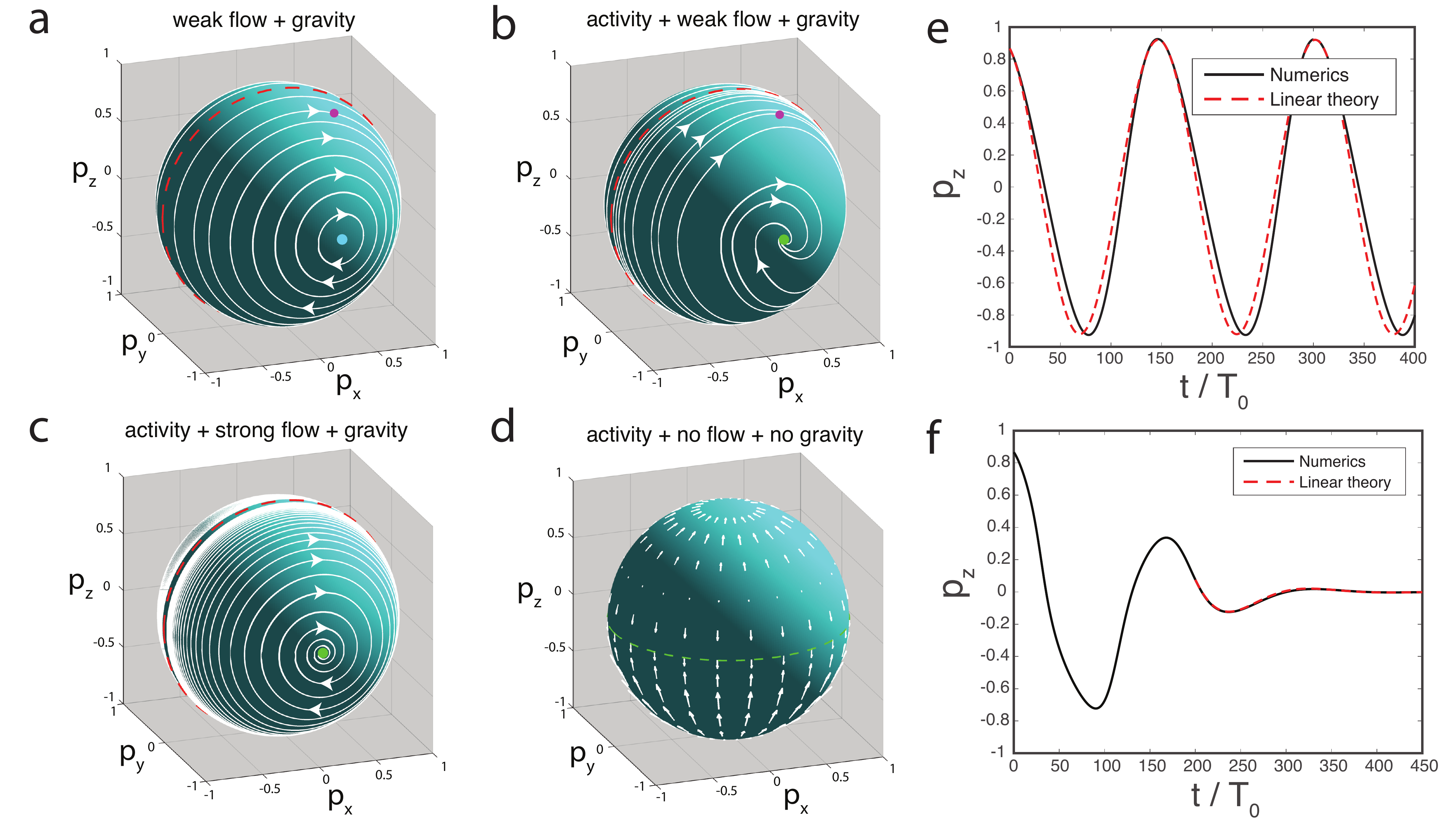}
\caption{\label{fig:phase_portraits} Phase portraits on the sphere $|\mathbf{p}| = 1$ for (a) an inactive, \textcolor{black}{bottom-heavy} particle in shear flow with $\dot{\gamma} R/U_{0} = 0.1$, and (b-c) for an active, \textcolor{black}{bottom-heavy} particle in a flow with (b) $\dot{\gamma} R/U_{0} = 0.1$ and (c) $\dot{\gamma} R/U_{0} = 0.5$.  The red dashed lines indicate the plane of mirror symmetry $p_{y} = 0$.  The cyan circle in (a) indicates the center of oscillatory motion. Green circles in (b) and (c) indicate stable fixed points (``attractors''). For (a) and (b), the magneta circles show the initial conditions for (a) the trajectory in Fig. \ref{fig:traj_pdfs_combined}(a) and (b) the trajectory in Fig. \ref{fig:traj_pdfs_combined}(b). The parameters characterizing the particle heaviness and activity are given in Methods. \textcolor{black}{(d) Vector field on the unit sphere representing the contribution of activity to the motion in (b) and (c). The green dashed line indicates $p_{z} = 0$.}  (e) Oscillation in $p_{z}$ with time for the inactive particle in Fig.  \ref{fig:traj_pdfs_combined}(a), also corresponding to the magenta circle in (a). The black dashed line gives the result obtained by numerical integration. The red dashed line gives the analytical solution for the linearized equations. The slight disagreement between them is due to the large amplitude of the oscillation, for which the effect of the nonlinearity in Eqs. \ref{eq:dynamical_eqns}-\ref{dot_pz} is important. (f) For the active particle in Fig. \ref{fig:traj_pdfs_combined}(b), with initial conditions corresponding to the magneta circle in (b), the oscillation in $p_{z}$ decays with time, and the orientation $\mathbf{p}$ eventually approaches the stable fixed point $\mathbf{p}^{*}$. The analytical solution is shown for the later part of the decay, for which $p_{z}$ has small amplitude and is well described by linear theory. }
\end{figure*}

\begin{equation}
\label{eq:effective_osc}
\ddot{\tilde{p}}_{z} = - \left(\frac{1}{2} \dot{\gamma} f(H/R)  p^{*}_{y} \right)^{2} \tilde{p}_{z} + \frac{\partial \Omega_{x'}^{\textcolor{black}{(ax)}}}{\partial p_{z}} \biggr\rvert_{p_{z}^{*} = 0, H} \dot{\tilde{p}}_{z}.
\end{equation}

\noindent \textcolor{black}{The motion of the orientation vector} resembles a damped harmonic oscillator with intrinsic frequency $\omega_{0} = \frac{1}{2} \dot{\gamma} f(H/R)  p^{*}_{x}$. \textcolor{black}{Recalling that inactive, bottom-heavy particle were observed to continuously rotate in the experiments, we consider the case in which $\Omega_{x'}^{\textcolor{black}{(ax)}} =  \Omega_{x'}^{(g)}$.}  We note that 

\begin{equation}
\label{eq:PartialOmegaPartialPz_Justgrav}
 \frac{\partial \Omega_{x'}^{(g)}}{\partial p_{z}} \biggr\rvert_{p_{z}^{*} = 0, H} \textcolor{black}{=  \frac{\partial \dot{\theta}^{(g)}}{\partial \theta} \biggr\rvert_{\theta^{*} = 90^{\circ}, H}} = 0,
\end{equation}

\noindent \textcolor{black}{where we have used $\dot{\theta}^{(ax)} = -\Omega_{x'}^{(ax)}$  and $\dot{\theta}^{(g)} \sim \sin(\theta)$.} Therefore, the dissipative term in Eq. \ref{eq:effective_osc} is zero, and the orientation $\mathbf{p}$ has a continuous family of closed orbits centered on $\mathbf{p}^{*}$.
Now we consider active, bottom-heavy particles, for which  $\Omega_{x'}^{\textcolor{black}{(ax)}} = \Omega_{x'}^{(s)} + \Omega_{x'}^{(g)}$. The fixed point is a stable attractor if 

\begin{equation}
\label{eq:stability_attractor}
 \frac{\partial \Omega_{x'}^{(s)}}{\partial p_{z}} \biggr\rvert_{p_{z}^{*} = 0, H}  \textcolor{black}{=  \frac{\partial \dot{\theta}^{(s)}}{\partial \theta} \biggr\rvert_{\theta^{*} = 90^{\circ}, H}} < 0.
\end{equation}

\noindent \textcolor{black}{This condition has the following interpretation: if the particle orientation is perturbed out of the plane of the wall, the active contribution to rotation responds (increases or decreases)  so as to oppose the perturbation.} \textcolor{black}{If this condition is satisfied,} activity \textcolor{black}{induces an effective ``friction''}  that damps oscillations of the orientation $\mathbf{p}$,  \textcolor{black}{driving attraction of $\mathbf{p}$ to $\mathbf{p}^{*}$. \textcolor{black}{Satisfaction of the condition depends on the details of the interactions between the particle and the wall that originate in swimming activity. In turn, these details depend on the character of the self-propulsion mechanism (see, e.g., the comparison of pushers and pullers in Ref. \cite{spagnolie12}.)}}



\subsubsection*{Application to a self-phoretic particle}
\textcolor{black}{Up to this point, we have proceeded without specifying a model for the particle composition and self-propulsion, and we derived and analyzed  Eqs. \ref{eq:dynamical_eqns}-\ref{dot_h} in general terms. Now we seek to obtain illustrative particle trajectories by numerical integration. Accordingly,  hereafter we calculate the various terms in Eqs. \ref{eq:dynamical_eqns}-\ref{dot_h} by using a simple, well-established model of neutral self-diffusiophoresis in confinement \cite{golestanian05, golestanian07, uspal15a, simmchen16, mozaffari16, ibrahim16}}. In the model, the particle has a hemispherical catalytic cap (Fig. \ref{fig:schematic}(b), black), and the orientation vector $\mathbf{p}$ points from its catalytic pole to its inert pole. The particle emits solute molecules (i.e., oxygen) at a constant, uniform rate from its cap, leading to self-generated solute gradients in the surrounding solution. These gradients drive surface flows on the particle (Fig. \ref{fig:schematic}(b), magenta arrows) and on the wall (blue arrows), leading to directed motion of the particle. \textcolor{black}{We note that our generic model does not account for effects (e.g., ionic or  electrokinetic) specific to the detailed self-phoretic mechanism \cite{brown14,ebbens2014,brown2016,brown2017}. Further details are provided in Methods}.

We consider some illustrative examples, using dimensionless parameters comparable to those in the experiments (see Methods for definition and estimation of parameters). We fix the particle height as $H/R = 1.2$.  In Fig. \ref{fig:traj_pdfs_combined}(a), we show a particle trajectory in a shear flow with strength $\dot{\gamma} R/U_{0} = 0.1$ and initial orientation $\theta_{0} = 30^{\circ}$ and $\phi_{0} = 315^{\circ}$. The particle rotates so that its inert face points largely in the $-\hat{y}$ direction, but with a small downstream orientation ($p_{x} > 0$). Notably, this slight downstream orientation agrees with experimental observations. If the particle is inactive (but still bottom-heavy), then from the same initial orientation, the particle simply translates in the flow direction (Fig. \ref{fig:traj_pdfs_combined}(b)), rotating as it does so. 

\textcolor{black}{Since $h = H$ is taken to be constant}, \textcolor{black}{the instantaneous value of} $\mathbf{p}$ \textcolor{black}{completely} determines the instantaneous velocity of the particle. In Fig. \ref{fig:phase_portraits}(a)-(c), phase space trajectories are shown on the unit sphere $|\mathbf{p}| = 1$. The initial orientation in Fig. \ref{fig:traj_pdfs_combined}(a) and (b) is indicated by magenta circles in Fig. \ref{fig:phase_portraits}(a) and Fig. \ref{fig:phase_portraits}(b), respectively. We see that an inactive particle (Fig. \ref{fig:phase_portraits}(a)) has a continuous family of closed orbits in $\mathbf{p}$. When the particle is active, these oscillations are damped, and trajectories in phase space are attracted to the fixed point (Fig. \ref{fig:phase_portraits}(b)-(c)). In (c), we show the effect on the structure of trajectories as the flow strength is increased. For stronger flows, the approach to the fixed point is more oscillatory. In Fig. \ref{fig:phase_portraits}(e), we plot the component $p_{z}$ (black line) for the inactive particle in Fig. \ref{fig:traj_pdfs_combined}(a). In Fig. \ref{fig:phase_portraits}(f), we show $p_{z}$ for the active particle in Fig. \ref{fig:traj_pdfs_combined}(b). We can also obtain analytical solutions to the linearized equations, characterized by a few input parameters that are evaluated numerically (see SM). These solutions are shown as dashed red lines in Figs. \ref{fig:phase_portraits}(e) and (f), and agree well with the numerical data. \textcolor{black}{Additionally, we note that if the particle height is allowed to change, one can still obtain the phenomenology studied here of attraction to a planar alignment steady state; an example is given in \textcolor{black}{SI Sect. S8}.}

\textcolor{black}{The phase portraits also provide an intuitive way to understand the stability condition expressed in Eq. \ref{eq:stability_attractor}. Consider the bottom-heavy, inactive particle with the phase portrait shown in Fig. \ref{fig:phase_portraits}(a). How does adding activity transform this portrait into that shown in Fig. \ref{fig:phase_portraits}(b)? The contribution of activity to $\dot{\mathbf{p}}$ is shown as the vector field in Fig. \ref{fig:phase_portraits}(d). Since this vector field is small but non-zero on the equator $p_{z} = 0$, adding it to the portrait in Fig. \ref{fig:phase_portraits}(a) will shift the center of oscillation (cyan circle) slightly towards the axis $p_{x} = 0$, producing the green circle in Fig. \ref{fig:phase_portraits}(b). More significantly, the addition of this vector field will destabilize continuous oscillatory motion in the following way. Consider the closed trajectories in Fig.  \ref{fig:phase_portraits}(a) that are closest to the cyan circle.  In the neighborhood of the equator, activity always ``pushes'' the vector $\mathbf{p}$ towards the equator  (here neglecting the small value on the equator, the main effect of which is to shift the fixed point.)   Therefore, over each period of oscillation, the orientation vector will get slightly closer to the fixed point -- transforming the closed circular orbits into decaying spiral orbits.}


\subsubsection*{Effect of thermal noise}

All of the preceding analysis assumed deterministic motion. \textcolor{black}{As a further exploration, we consider} the effect of thermal noise on the particle orientation by performing Brownian dynamics simulations (see Methods). In Fig. \ref{fig:traj_pdfs_combined}(c), we show the probability distribution functions for the $\phi$ obtained for a Janus particle driven by shear flow at dimensionless inverse temperature $Pe_{p} = 500$ (see Methods for definition) and shear rate $\dot{\gamma}R/U_{0} = 0.1$. The distribution is symmetric, and has two peaks near the \textcolor{black}{steady angle} $\phi^{*} = \pm \textcolor{black}{80.9}^{\circ}$ \textcolor{black}{predicted by the deterministic model} (Fig. \ref{fig:traj_pdfs_combined}(c)). For a higher shear rate $\dot{\gamma}R/U_{0} = 0.5$, we find that the peaks in $\phi$ are sharper (Fig. \ref{fig:traj_pdfs_combined}(d)), in qualitative agreement with the experiments, \textcolor{black}{with the peaks shifted near the deterministic prediction for this shear rate, $\phi^{*} = \pm {88.2}^{\circ}$.} Additional probability distributions for the components of $\mathbf{p}$ are given in \textcolor{black}{Fig. S14}.  


\section{Discussion}
In this paper, \textcolor{black}{we use catalytic Janus particles as a well-controlled model experimental system to study spherical active particles in confined flows.} We demonstrate that \textcolor{black}{spherical} active particles near surfaces, when exposed to external flows, \textcolor{black}{can} exhibit robust alignment and motion along the cross-stream direction. Our model reveals how this behaviour arises from the interplay of shear flow and swimming in confinement. \textcolor{black}{The steady orientation is determined by a balance of contributions to the angular velocity of the particle from shear flow, bottom-heaviness, and swimming near a planar substrate.
Near-surface swimming introduces an effective ``friction''} opposing rotation of the particle away from the preferred orientation. \textcolor{black}{The mechanism is generic in the sense that it can occur for any spherical microswimmer with axisymmetric actuation, and is not specific to a particular mechanism of propulsion (e.g., chemical or mechanical.)}  \textcolor{black}{As a consequence of the alignment, the particles migrate across the streamlines of the external flow as they are carried downstream.}

To the best of our knowledge, our results are the first to demonstrate that swimmer/surface interactions (e.g. hydrodynamic interactions) can drive a rich directional response of spherical particles to external flows. This is in contrast with previous works on natural microswimmers (e.g., bacteria) where complex body shapes and flagellar beat patterns were implicated in directional response \cite{marcos12, bukatin15}. Additionally, we have obtained semi-quantitative agreement between particle orientational statistics obtained from the experiments and from the theoretical model.
 For lab-on-a-chip devices that use continuous flows and artificial microswimmers, our findings imply that the microswimmers would have a tendency to migrate to the confining side walls of the device. Our findings additionally raise the possibility that in dense suspensions of microswimmers, for which fluid flows are self-generated, the collective behavior of the suspension may be sensitive to the detailed interactions between individual microswimmers and bounding surfaces.

\section{Materials and methods}
\subsubsection*{Sample preparation}
The Janus particles were obtained by electron-beam deposition of a Pt layer on a monolayer of silica colloids. The monolayers were prepared either by a Langmuir-Blodgett (LB) method or a drop casting method. For the LB method, silica colloids ($R = 1 \, \mu m$, Sigma Aldrich) were first surface treated with allyltrimethoxysilane to make them amphiphilic. A suspension of these particles in chloroform-ethanol mixture (80/20 $v/v$) was carefully dropped onto the LB trough and compressed to create a closely packed monolayer. The monolayer was then transferred onto a silicon wafer at a surface pressure of $20 \, mN/m$.  The silicon wafer was then shifted to a vacuum system for the electron beam deposition of a thin layer of Pt ($10 \, nm$) at $10^{-6} \, mmHg$. The Janus particles were released into deionized water using short ultrasound pulses. The suspension of Janus particles in water was stored at room temperature. The monolayers of $R = 2.5 \, \mu m$ silica particles (Sigma Aldrich) were prepared by drop casting of the suspension of colloids onto an oxygen plasma treated glass slide. The plasma treatment was used to make the glass slide hydrophilic and ensure uniform spreading of the particle suspension. The solvent was subsequently removed by slow evaporation. The Pt deposition step was identical to the one used for $R = 1 \, \mu m$ particles. 

\subsubsection*{Tracking}
Particle tracking was performed using an automated tracking program developed in-house. The Python based program uses OpenCV library for image processing and numpy for data handling. In source videos filmed in grayscale, each frame is first cleaned of noise by using blurring techniques, which substitute each pixel with an average of its surroundings. The particles are then separated from the background by using either of the two segmentation methods: threshold and gradient. In the threshold method, given a grayscale image \textit{img(x,y)}, and a threshold value \textit{T}, this operation results in a binary image \textit{out(x,y)} given by:

\begin{equation}
\text{out}(x,y) = 
\left\{
\begin{array}{cc}
1  & \text{if img}(x,y) \geq T \\
0 & \text{if img}(x,y) < T
\end{array}
\right.
\end{equation}

\noindent The gradient method is used for images with irregular brightness or when the particles are hard to distinguish from the background. In the first step of this method the gradient of the image \(\nabla img = (\frac{\partial img}{\partial x}, \frac{\partial img}{\partial y}) \), is approximated by convolving the original frame with a Sobel operator. This results in two images, one which is the derivative along the X axis and another along the Y axis. These images are then thresholded and joined together to obtain the segmented image. The final result has the edges of the detected particles.

The center of each particle is approximated as the center of mass of the contours obtained after segmentation. Particle trajectories are calculated using Bayesian decision-making, linking every particle center with the previous closest one. Intermediate missing positions, if any, are interpolated using cubic or linear splines. \\
In order to calculate the $\mathbf{p}$  vector, a line of predefined radius (approximately equal to the particle radius) is drawn through the center at a test angle. The standard deviation of the pixel values along the test-line is calculated and stored. The process is repeated multiple times and the standard deviations are compared. The test-angle with the least standard deviation is assumed to correspond to the separator between the silica and Pt halves and the vector orthogonal to it, the orientation vector (see Fig. S2).

\subsubsection*{Theoretical calculation of particle velocity}

Here, we present our model for calculation of the particle velocities $\mathbf{U}$ and $\mathbf{\Omega}$ as a function of $h$ and $\mathbf{p}$.  We take the instantaneous position of the particle to be $\mathbf{x}_{p} = (x_{p}, y_{p}, h)$ in a stationary reference frame.  The catalytic cap emits a product molecule at a constant and uniform rate $\kappa$. We take the solute number density $c(\mathbf{x})$ to be quasi-static, i.e., it obeys the equation $\nabla^{2} c = 0$ with boundary conditions $-D \hat{n} \cdot \nabla c = \kappa$ on the cap, $\hat{n} \cdot \nabla c  = 0$ on the inert region of the particle surface, and $\hat{n} \cdot \nabla c = 0$ on the wall. Here, $D$ is the diffusion coefficient of the solute molecule, $\mathbf{x}$ is a location in the fluid, $\hat{n}$ and is the normal vector pointing from a surface into the liquid. For each instantaneous configuration ($h$, $\mathbf{p})$, this set of equations can be solved for $c(\mathbf{x})$, e.g., numerically by using the boundary element method \cite{pozrikidis02,uspal15b}.  

We take the velocity $\mathbf{u}(\mathbf{x})$ in the fluid solution to obey the Stokes equation $-\nabla P + \eta \nabla^{2} \mathbf{u} = 0$, where $\eta$ is the solution viscosity and $P(\mathbf{x})$ is the pressure in the solution. Additionally, the fluid is incompressible, so that $\nabla \cdot \mathbf{u} = 0$. The Stokes equation is a linear equation. Therefore, the contributions to $\mathbf{U}$ and $\mathbf{\Omega}$ from various boundary conditions for $\mathbf{u}$, as well as from various external forces and torques, can be calculated individually as the solution to separate subproblems and then superposed. Hence, we write $\mathbf{U} = \mathbf{U}^{(f)} + \mathbf{U}^{\textcolor{black}{(ax)}}$ and $\mathbf{\Omega} = \mathbf{\Omega}^{(f)} + \mathbf{\Omega}^{\textcolor{black}{(ax)}}$, where $(f)$ indicates the contributions of the external \textit{\textbf{f}}low, and the \textit{\textbf{a}}xi\textit{\textbf{s}}ymmetric contributions are from \textit{\textbf{g}}ravity and \textcolor{black}{\textbf{\textit{s}}}wimming activity: $\mathbf{U}^{\textcolor{black}{(ax)}} = \mathbf{U}^{(g)} + \mathbf{U}^{\textcolor{black}{(s)}}$ and $\mathbf{\Omega}^{\textcolor{black}{(ax)}} = \mathbf{\Omega}^{(g)} + \mathbf{\Omega}^{\textcolor{black}{(s)}}$. For each subproblem, the fluid is governed by the Stokes equation and incompressibility condition.

We first consider the subproblem for the contribution of the external flow. The fluid velocity is subject to no-slip boundary conditions $\mathbf{u} = 0$ on the planar wall and $\mathbf{u} = \mathbf{u}_{ext} + \mathbf{U}^{(f)} + \mathbf{\Omega}^{(f)} \times (\mathbf{x} - \mathbf{x}_{p})$ on the particle. Here, $\mathbf{u}_{ext}$ is the external flow velocity, $\mathbf{u}_{ext} = \dot{\gamma} z \hat{x}$. Additionally, the particle is free of external forces and torques, closing the system of equations for $\mathbf{U}^{(f)}$ and $\mathbf{\Omega}^{(f)}$. The solution of this subproblem is well-known; see, for instance, Goldman et al. \cite{goldman67}
 
The two axisymmetric subproblems are calculated in the ``primed'' frame co-rotating with the particle (Fig. \ref{fig:schematic}(b)). For the subproblem associated with particle activity, we employ the classical framework of neutral self-diffusiophoresis \cite{anderson89, golestanian05,golestanian07,uspal15a,simmchen16}. In this subproblem, the self-generated solute gradients drive surface flows on the wall and the particle surface, $\mathbf{v}_{s} = -b(\mathbf{x}_{s}') \nabla_{||} c(\mathbf{x}')$, where $\nabla_{||} \equiv (\mathbf{1} - \hat{n}\hat{n}) \cdot \nabla$ and $\mathbf{x}_{s}'$ is a location on a surface (in the primed frame). The ``surface mobility'' $b(\mathbf{x}_{s}')$ encapsulates the details of the molecular interaction between the solute and the bounding surfaces. We  write the boundary conditions $\mathbf{u} = \mathbf{U}^{\textcolor{black}{(s)}} + \mathbf{\Omega}^{\textcolor{black}{(s)}} \times (\mathbf{x}' - \mathbf{x}_{p}') + \mathbf{v}_{s}(\mathbf{x}_{s}')$ on the particle, and $\mathbf{u} = \mathbf{v}_{s}(\mathbf{x}_{s}')$ on the wall. Again, specifying that the particle is force and torque free closes the system of equations for $\mathbf{U}^{\textcolor{black}{(s)}}$ and $\mathbf{\Omega}^{\textcolor{black}{(s)}}$. This subproblem can be solved numerically using the boundary element method \cite{pozrikidis02,uspal15b}. For the subproblem associated with gravity, we use the ``eggshell'' model of Campbell and Ebbens for the shape of the cap, taking the cap thickness to vary smoothly from zero at the particle ``equator'' to a maximum thickness of $t$ at the active pole \cite{Ebbens2013}. Details concerning this subproblem, which is solved using standard methods \cite{simmchen16}, are provided in Supplementary Materials.


We now specify the parameters characterizing the system.  We choose to take the inert and catalytic regions 
of the particle to have different surface mobilities, $b_{inert}$ and $b_{cap}$, with $b_{inert}/b_{cap} = 
0.3$ and $b_{cap} < 0$. For this parameter, a neutrally buoyant Janus particle, when it is far away from bounding surfaces, moves in the $\mathbf{p}$ direction (i.e., away from its cap) with a velocity $U_{fs} = 13/80 \; U_{0}$, where $U_{0} = |b_{cap}| \kappa/D$ \cite{michelin14}. The wall is characterized by a surface mobility $b_{w}$. We choose $b_{w}/b_{cap} = -0.35$. These surface mobility ratios are chosen to be similar to those used in previous work (Ref. \cite{simmchen16}, where we had $b_{inert}/b_{cap} = 0.3$ and $b_{w}/b_{cap} = -0.2$), and to give a slightly downstream steady orientation. We non-dimensionalize length with $R$, velocity with $U_{0}$, and time with $T_{0} = R/U_{0}$. In order to non-dimensionalize the gravitational and shear contributions, we must estimate $U_{0}$ in real, dimensional units. Rather than calculate $U_{0}$ directly, which requires estimates for $\kappa$ and $|b_{cap}|$, we use the expression $U_{fs} = 13/80 \; U_{0}$ \cite{michelin14}. Knowing that, experimentally, the particle characteristically moves at $U_{fs} \approx 5\;{\mu m/s}$ when it is far from surfaces, we obtain $U_{0} \approx 30\;{\mu m/s}$. The parameters describing the heaviness of the particle are given in Supplementary Information.  
Finally, we consider the shear rate $\dot{\gamma}$. This is not known experimentally, but can be roughly estimated by considering inactive particles to act as passive tracers. From Goldman et al., the velocity of a spherical particle driven by shear flow near a wall is $U^{(f)}_{x} = g(h/R) \dot{\gamma} h$ \cite{goldman67}. Experimentally, inactive particles in flow are observed to move with $V^{*} \sim 10 \, {\mu m/s}$. The particle height $h$ is difficult to observe experimentally, but we take it to be set by the balance of gravity and electrostatic forces. Hence, the particle/wall gap $\delta$ is on the order of a Debye length $\lambda_{D} \sim 0.1 R$, so that $h \sim 1.1R$. For $h/R \geq 1.05$, the factor $g(h/R) \approx 1$. We therefore estimate $\dot{\gamma} \approx 10\, {s^{-1}}$, and a typical dimensionless shear rate to be $\dot{\gamma} R/U_{0} \approx 0.8$.  As a reminder, our aim is to establish semi-quantitative agreement with experiments, and therefore we seek only order of magnitude accuracy in the dimensionless parameters.
 
In assuming the concentration field to be quasi-static, we neglected the advective effects on the solute field by the external shear flow and by the finite velocity of the particle. These approximations are valid for small Peclet numbers $Pe \equiv U_{fs} R/D$ and $Pe_{\dot{\gamma}} \equiv \dot{\gamma} R^{2}/D$. At room temperature, the diffusion coefficient of oxygen is $D \sim 4 \times 10^{-9} m^{2}/s$ \cite{popescu10}, so that $Pe \approx 0.003$ and $Pe_{\dot{\gamma}} \approx 0.015$. Furthermore, in taking the fluid velocity $\mathbf{u}$ to be governed by the Stokes equation, we neglected fluid inertia. This approximation is justified for low Reynolds number, $Re = \rho_{fluid} U_{0} R/ \eta$, where $\eta$ is the dynamic viscosity of the solution. Using $\eta \sim 10^{-3} Pa\,s$ for water, we obtain $Re \approx 10^{-4}$. 


\subsubsection*{Effect of thermal noise}

The particle Peclet number $Pe_{p} = U_{fs} R/D_{0}^{t}$ characterizes the relative strengths of self-propulsion and translational diffusion of the particle. Here, $D_{0}^{t} = k_{B} T/6 \pi \eta R$ is the translational diffusion coefficient of the particle in free space. For a particle with $R = 2.5\;{\mu m}$ in water  at room temperature ($k_{B}T_{room} \sim 4 \times 10^{-21} J$) with $U_{fs} \approx 5\;{\mu m\textcolor{black}{/s}}$, we estimate $Pe_{p} \approx 880$.  In order to perform Brownian dynamics simulations, we adapt the Euler-Maruyama integration scheme introduced by Jones and Alavi \cite{Jones92}, and later presented by Lisicki \textit{et al.} \cite{Lisicki14}, which explicitly includes the effects of the wall on diffusion. Our principal modification to the method is inclusion of deterministic contributions to $\dot{\mathbf{p}}$ (Eqs. \ref{dot_px}-\ref{dot_pz}). Further details are provided in Supplementary Information and Ref. \cite{Lisicki14}. 

\acknowledgments
The authors thank M. N. Popescu for  helpful discussions. W.E.U. acknowledges financial support from the DFG, grant no. TA 
959/1-1. S.S., J.S. and J.K. acknowledge the DFG grant no. S.A 2525/1-1. S.S. acknowledges the Spanish MINECO for grant CTQ2015-68879-R (MICRODIA). This research also has received funding from the European Research Council under the European Union's Seventh Framework Programme (FP7/2007-2013)/ERC grant agreement 311529. 

\bibliography{janus_shear}




\section*{Supplementary material}

# Supplementary Material for "Cross-stream migration of active particles"

Jaideep Katuri,[1,2] William E. Uspal,[1,3,*] Juliane Simmchen,[1] Albert Miguel-Lopez,[2] and Samuel Sanchez[1,2,4,†]

[1] Max-Planck-Institut für Intelligente Systeme, Heisenbergstr. 3, 70569 Stuttgart, Germany
[2] Institut de Bioenginyeria de Catalunya (IBEC), Baldiri I Reixac 10-12, 08028 Barcelona, Spain
[3] IV. Institut für Theoretische Physik, Universität Stuttgart,
Pfaffenwaldring 57, D-70569 Stuttgart, Germany
[4] Institució Catalana de Recerca i Estudis Avancats (ICREA),
Pg. Lluís Companys 23, 08010, Barcelona, Spain
(Dated: December 7, 2017)

[*] uspal@is.mpg.de
[†] ssanchez@ibecbarcelona.eu

## S1. ACTIVITY INDUCED RE-ORIENTATION OF PARTICLES

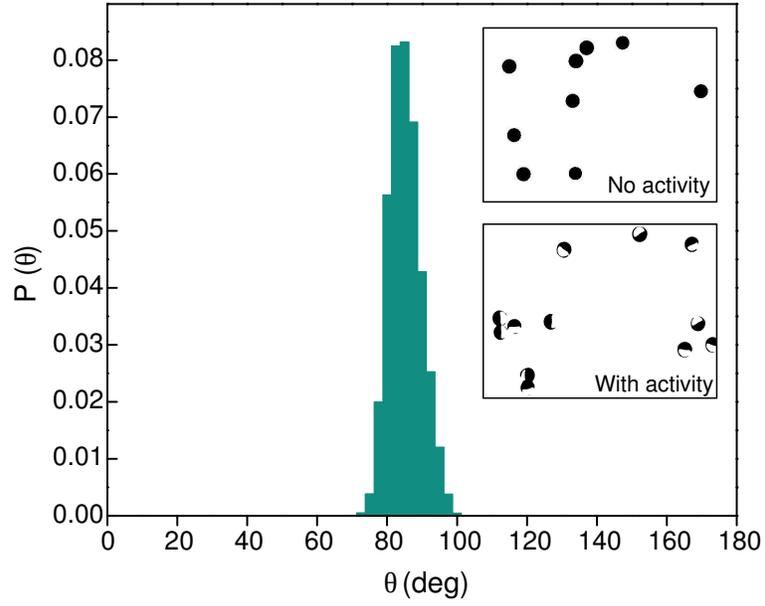

FIG. S1. Experimentally measured probability distribution of $\theta$ for active particles. Inset: Snapshots showing the system of Janus particles without and with activity. Activity changes the bottom heaviness induced $\theta = 0°$ to $\theta = 90°$.

## S2. METHOD TO DETECT THE IN-PLANE ANGLE OF JANUS PARTICLES

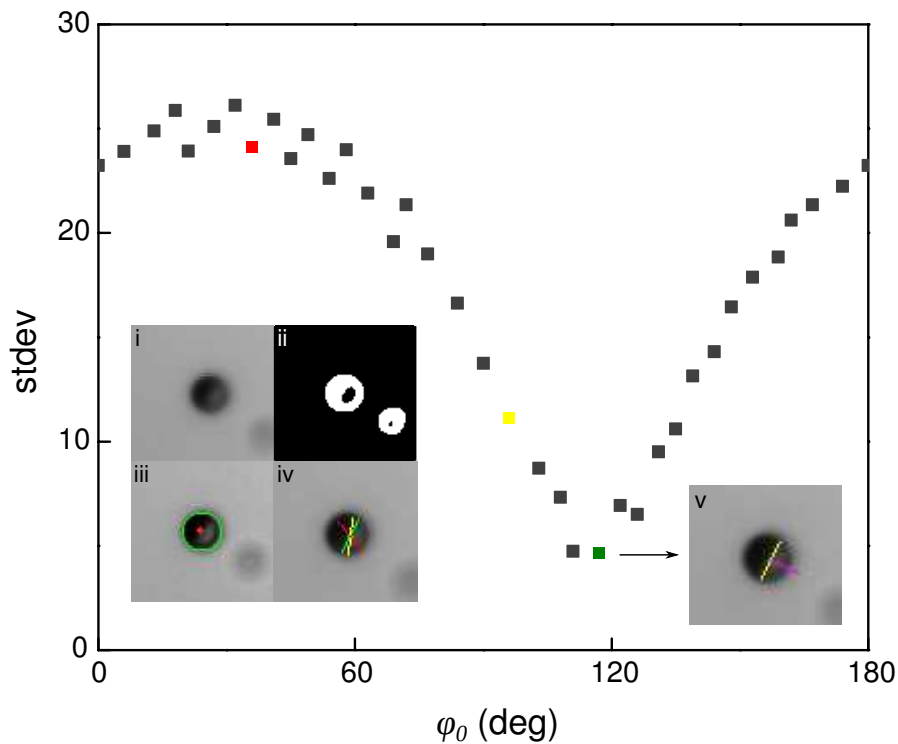

FIG. S2. The plot shows the standard deviation of the gray scale values along different lines ($l = 2r$) crossing the particle through the center. Each line is defined by an angle ($\phi_0$), calculated from the reference system of the image. The angle with lowest standard deviation corresponds to the direction that separates the black and white halves of the Janus particle. The insets show the process of center and angle detection. i) Original image, ii) binary image that results from applying adaptive thresholding, iii) Contour taken from the binary image, and center computed from the center of mass, iv) Different test lines used for the gray scale standard deviation calculation. Each one corresponds to a similarly colored point in the graph, v) The line with lowest standard deviation is selected and the final orientation ($\phi$) is taken to be the one perpendicular to that (pink), pointing towards the bright half of the Janus particle.

## S3. FLOW PROFILE IN SQUARE CAPILLARY

To examine the structure of the external flow **v**, we use the well-known series solution for viscous flow in a rectangular capillary [46]:

$$v_x(y,z) = v_0 \sum_{n,odd}^{\infty} \frac{1}{n^3} \left[ 1 - \frac{\cosh\left(n\pi \frac{y}{h_c}\right)}{\cosh\left(n\pi \frac{w_c}{2h_c}\right)} \right] \sin\left(n\pi \frac{z}{h_c}\right) \quad (1)$$

Here, $w_c$ is the width of the channel, and $h_c$ is the height of the channel, with $y \in [-w_c/2, w_c/2]$ and $z \in [0, h_c]$. The parameter $v_0$ characterizes the strength of the flow. The velocity components $v_y$ and $v_z$ vanish everywhere. For $w_c = h_c = 1000~\mu m$, the theoretical flow profile is shown in Fig. S3(a), scaled by the maximum value of the flow speed. The series was truncated at $n = 50$.

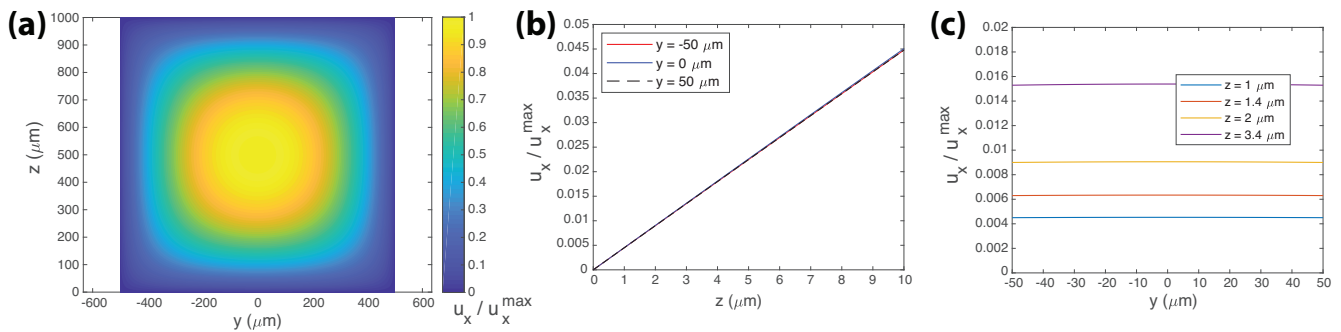

FIG. S3. (a) Speed of the flow, scaled by the maximum value, as a function of the position in the cross-section of the channel. (b) Speed of the flow, scaled by the maximum value, as a function of the vertical position $z$ in the vicinity of $z = 0~\mu m$, for several fixed values of $y$ near $y = 0~\mu m$. (c) Speed of the flow, scaled by the maximum value, as a function of the lateral position $y$ in the vicinity of $y = 0~\mu m$, for several fixed values of $z$ near $z = 0~\mu m$.

In order to determine the validity of the linear shear flow approximation, we focus on a spatial window near $y = 0$ and $z = 0$, which is the center of the bottom wall of the channel. In Fig. S3(b), we show how the flow speed varies in the $z$ direction for several values of $y$ in the vicinity of $y = 0~\mu m$. In the region $y \in [-50~\mu m, 50~\mu m]$, the flow can clearly be approximated as increasing linearly in the $z$ direction, with negligible variation of the shear rate (i.e., the slope of the flow profile) with $y$. As a second check, in Fig. S3(c), we examine how the flow speed changes with $y$ for several fixed values of $z$ near $z = 0~\mu m$. We can clearly neglect variation of the flow speed in the $y$ direction.

## S4. ESTIMATION OF SHEAR RATE FROM PARTICLE ROTATION

Fig. 2a in the main text, repeated as Fig. S4 here, shows the experimentally observed rotation time $\tau$ as a function of the observed particle speed $V^*$ for passive (inactive) Janus spheres driven by controlled external flow. Our aim in this section is to extract the particle height $h_p$ (where "p" stands for "passive") from the data, as well as the shear rate $\dot{\gamma}$ as a function of $V^*$, thereby obtaining $\dot{\gamma}(V^*)$.

To do so, we develop a model for a bottom-heavy Janus sphere driven by shear flow over a planar wall. We assume that the sphere is initially allowed to relax to thermal equilibrium in quiescent fluid. The flow is turned on at time $t = 0$. Therefore, at $t = 0$, the particle has an initial cap-down orientation. Subsequently, the particle rotates around an axis that we assume is parallel to the vorticity axis $\hat{y}$. Therefore, **p** always remains in the $xz$ plane. We define $\xi$ to be the angle **p** makes with the $\hat{z}$ direction, with $\xi \in [0, 2\pi]$. Additionally, the particle translates downstream in the $\hat{x}$ direction.

The angular velocity of the particle has two contributions. Bottom-heaviness tends to align the particle orientation vector **p** with the vertical:

$$\boldsymbol{\Omega}^{(g)} = B\,\mathbf{p} \times \hat{z} = -B\sin(\xi)\,\hat{y}. \quad (2)$$

Here, the parameter $B$ characterizes the strength of bottom-heaviness.

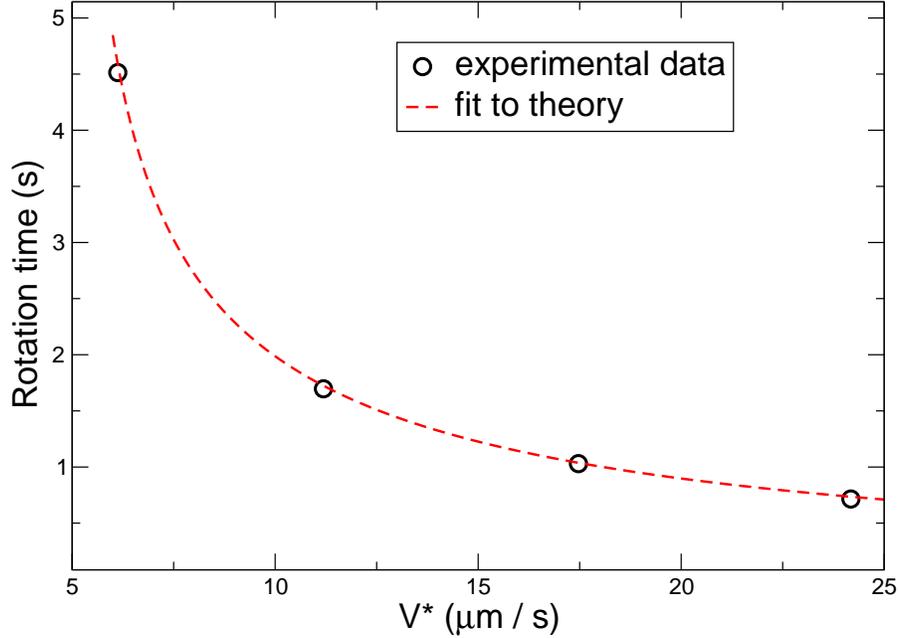

FIG. S4. Rotation time $\tau$ as a function of speed $V^*$ for inactive Janus particles in flow. The red dashed curve represents a fit to the theoretical function $\tau = \frac{2\pi}{\sqrt{aV^{*,2}+b}}$.

Secondly, shear flow drives rotation of the particle around the vorticity axis:

$$\mathbf{\Omega}^{(f)} = \frac{\dot{\gamma}}{2} f(h_p/R)\,\hat{y}. \tag{3}$$

Here, $h_p$ is the height of the particle above the planar surface. We take the height to be a constant quantity that is unaffected by shear flow. As discussed in detail in Section S6, $f(h/R)$ is a function representing hydrodynamic friction with the wall. The particle translates downstream with a velocity

$$\mathbf{U}^{(f)} = \dot{\gamma} h_p g(h_p/R)\,\hat{x}. \tag{4}$$

Likewise, as discussed in Section S6, the function $g(h/R)$ represents the effect of hydrodynamic friction from the wall. Therefore, the observed velocity is

$$V^* = \dot{\gamma} h_p g(h_p/R). \tag{5}$$

Noting that $\Omega_y = \dot{\xi}$, we write

$$\dot{\xi} = \frac{d\xi}{dt} = -B\sin(\xi) + \frac{\dot{\gamma}}{2} f(h_p/R). \tag{6}$$

For convenience, we define $C$ as the second term on the right hand side:

$$\dot{\xi} = \frac{d\xi}{dt} = -B\sin(\xi) + C. \tag{7}$$

We can separate $\xi$ and $t$:

$$\frac{d\xi}{-B\sin(\xi) + C} = dt \tag{8}$$

We know that $C > 0$, since shear is able to overcome the effect of bottom-heaviness and completely rotate the particle. We integrate the left hand size from $\xi = 0$ to $\xi = 2\pi$, and the right hand side from $t = 0$ to $t = \tau$, obtaining:

$$\tau = \frac{2\pi}{\sqrt{C^2 - B^2}}. \tag{9}$$

Now we recall that $C = \frac{\dot{\gamma}}{2} f(h_p/R)$. We can eliminate $\dot{\gamma}$ by using $V^* = \dot{\gamma} h_p g(h_p/R)$:

$$\tau = \frac{2\pi}{\sqrt{\left(\frac{f(h_p/R)}{2h_p g(h_p/R)}\right)^2 V^{*,2} - B^2}}. \tag{10}$$

Interestingly, this function diverges at a non-zero $V^* = V_c^*$. This represents the lower critical flow speed needed to overcome the effect of bottom-heaviness. Finally, the parameter $B$ can be calculated from the gravitational torque:

$$B = \frac{\tau_{max}^{(g)}}{\zeta^{r,||}(h_p/R)}. \tag{11}$$

Here, $\tau_{max}^{(g)}$ is the maximum value of the torque from bottom-heaviness, which is a material and geometry dependent parameter. The function $\zeta^{r,||}(h_p/R)$ is a hydrodynamic friction coefficient for rotation around an axis parallel to the wall.

Now we are in a position to understand how information can be extracted from the experimental data in Fig. 2a of the main text. We fit a function of the form

$$\tau = \frac{2\pi}{\sqrt{aV^{*,2} + b}}. \tag{12}$$

to the data. The fitting parameter $a$ can be used to find the height $h_p$, since the functions $f(h_p/R)$ and $g(h_p/R)$ can be calculated numerically, as described in Section S6. Fitting for $a$, we estimate that $h_p/R \approx 1.4$. We can combine this estimate with Eq. 5 to obtain $\dot{\gamma}$ as a function of $V^*$:

$$\dot{\gamma} \approx 1.22 \, V^*/R. \tag{13}$$

Our fit for $a$ is robust, because this parameter determines $\tau$ in the asymptotic limit of large $V^*$, where most of our experimental data is located.

7## S5. INTERIM STATES IN CROSS-STREAM MIGRATION OF ACTIVE PARTICLES

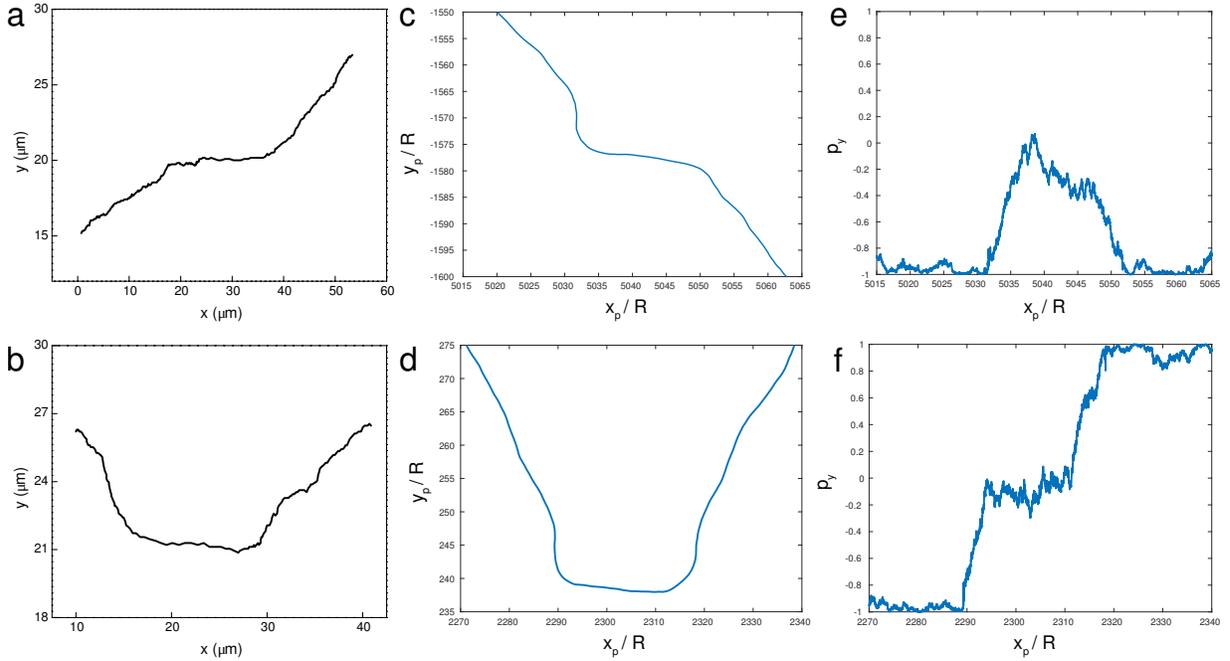

FIG. S5. Experimental (a-b) and numerically (c-d) obtained trajectories of active particles in a flow: interim states, reminisicent of bacterial "tumbling," can occur for cross stream migrating active particles when the particle orientation is against or with the flow ($\phi = 0°$ or $\phi = 180°$, i.e., $p_y = 0$.) For the numerically obtained trajectories in (c) and (d), $p_y$ is shown in (e) and (f), respectively. Eventually particles recover the cross stream behavior, either in the same direction as before beginning to "tumble," or the opposite one. Therefore, the interim states are associated with stochastic switching of the particle orientation between two bistable states with $p_y \approx \pm 1$.

## S6. EFFECT OF PARTICLE SIZE ON ORIENTATIONAL STABILITY

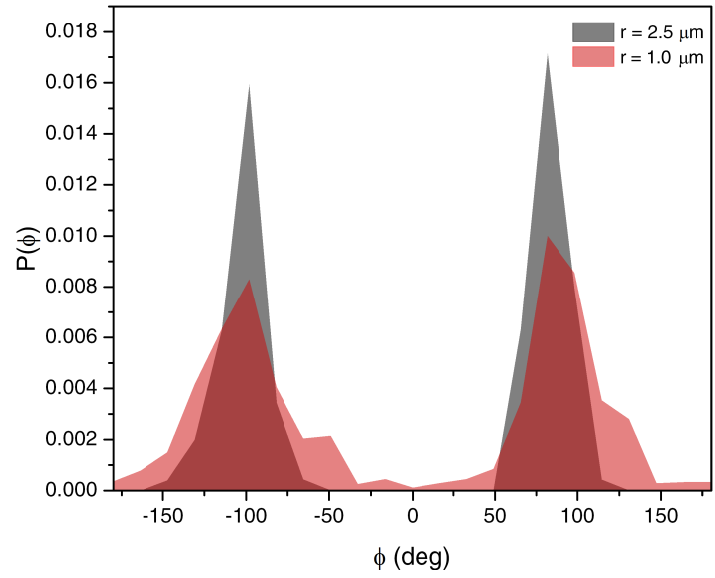

FIG. S6. Probability distribution function of $\phi$ for particles of $R = 1\ \mu m$ and $R = 2.5\ \mu m$ at $V_p \approx 6\ \mu m/s$ and $V^* \approx 24\ \mu m/s$.

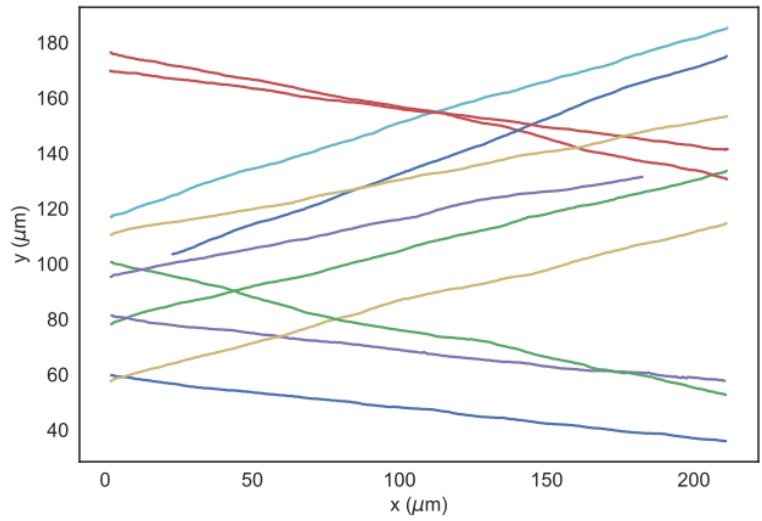

FIG. S7. Sample trajectories of $R = 2.5\ \mu m$ particles show very little deviation from their preferred cross-stream orientation at $V_p \approx 6\ \mu m/s$ and $V^* \approx 24\ \mu m/s$.

## S7. EFFECT OF PROPULSION VELOCITY ON ORIENTATIONAL STABILITY

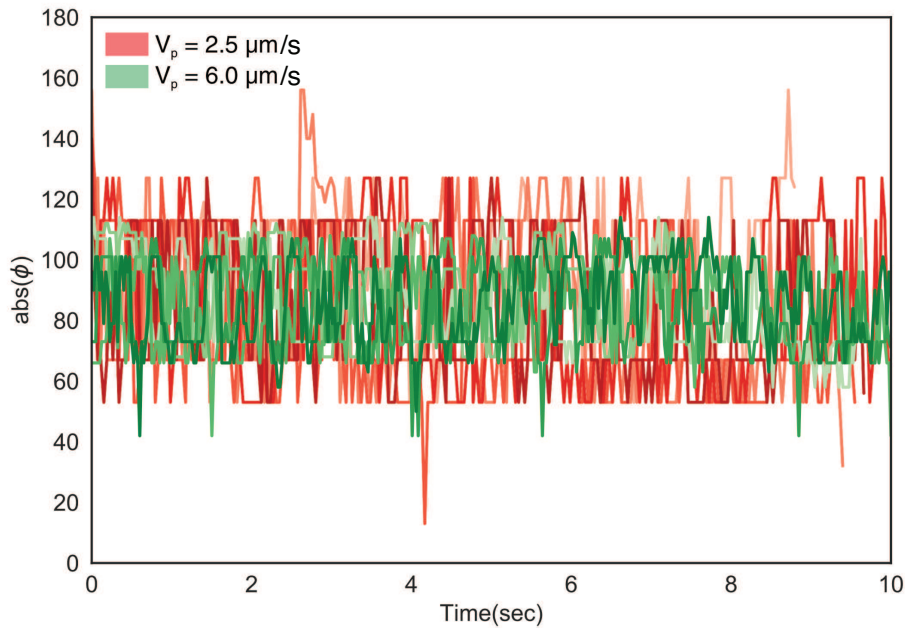

FIG. S8. Fluctuations in $\phi$ obtained from five different particles for two values of $V_p$ at $V^* \approx 24$ $\mu m/s$.

## S8. CONTRIBUTIONS OF SHEAR TO THE EQUATIONS OF MOTION

The functions $f(h/R)$ and $g(h/R)$ encode the influence of a planar surface on the motion of a spherical particle driven by shear. Friction from the surface leads to slower rotation and translation of the particle than would be obtained for an unconfined shear-driven particle. In particular, the functions $f(h/R)$ and $g(h/R)$ are the ratios of the near-wall rotational and translational velocities of the particle, respectively, relative to the bulk values $\Omega = \frac{1}{2}\dot{\gamma}$ and $U = \dot{\gamma}h$. These particle velocities would be obtained for a sphere in the same external flow profile $\mathbf{u} = \dot{\gamma}z\hat{x}$ without the presence of the wall. The two functions asymptote to one as the distance from the wall $h/R$ increases.

These functions were evaluated at selected values of $h/R$ by Goldman et al. by solving the incompressible Stokes equations in bispherical coordinates [47]. We obtain them numerically, for a larger number of values of $h/R$, by using the boundary element method (BEM). For other values of $h/R$, we use linear interpolation to obtain $f(h/R)$ and $g(h/R)$.

A direct comparison between numerical values of $f(h/R)$ and $g(h/R)$ obtained by Goldman et al. and from the BEM is shown in Table S1. Additionally, the full set of data obtained from the BEM is plotted in Fig. S9.

| $h/R$ | $f(h/R)$, Goldman et al. | $f(h/R)$, BEM | $g(h/R)$, Goldman et al. | $g(h/R)$, BEM |
|---|---|---|---|---|
| 1.0453 | 0.67462 | 0.67632 | 0.65375 | 0.65223 |
| 1.1276 | 0.77916 | 0.77923 | 0.76692 | 0.76429 |
| 1.5431 | 0.92368 | 0.92371 | 0.92185 | 0.91875 |
| 2.3524 | 0.97780 | 0.97784 | 0.97768 | 0.97439 |
| 3.7622 | 0.99430 | 0.99436 | 0.99436 | 0.99099 |

TABLE S1. Comparison of $f(h/R)$ and $g(h/R)$ as calculated by Goldman et al. and in this work using the boundary element method (BEM).

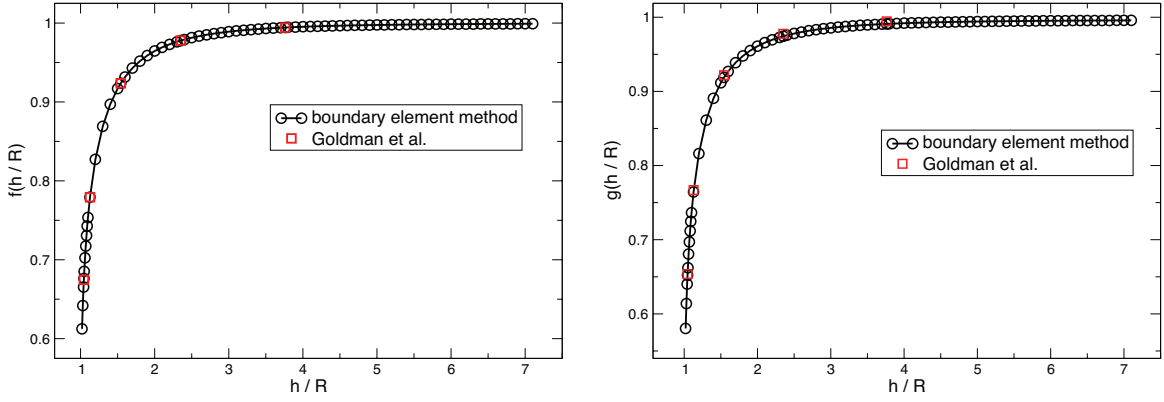

FIG. S9. The functions $f(h/R)$ and $g(h/R)$ as obtained by Goldman et al. and with the boundary element method. For a spherical particle driven by shear flow near a planar wall, the functions $f(h/R)$ and $g(h/R)$ are the ratios of the rotational and translational velocities, respectively, relative to the bulk values $\Omega = \frac{1}{2}\dot{\gamma}$ and $U = \dot{\gamma}h$. These velocities would be obtained for the same flow profile $\mathbf{u} = \dot{\gamma}z\hat{x}$ without the presence of a wall.

## S9. FIXED POINTS OF GOVERNING EQUATIONS

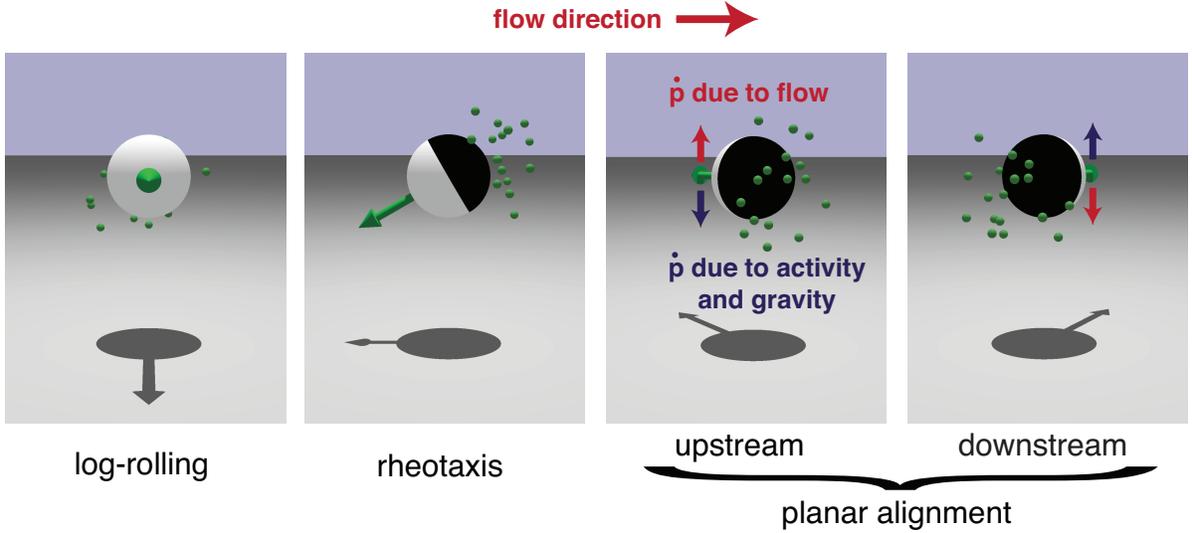

FIG. S10. Schematic illustration of the three fixed point solutions to Eqs. 2-5 in the main text. For a particle in a fixed point configuration $(\mathbf{p}^*, h^*/R)$, the time derivatives $\dot{\mathbf{p}} = 0$ and $\dot{h} = 0$, so that the particle translates in the $xy$ plane with a steady height and orientation. We focus on the *planar alignment* state, where $p_z^* = 0$, but the particle orientation vector has non-zero components $p_x^*$ and $p_y^*$ in the flow and vorticity directions, respectively. For planar alignment, the component $p_x^*$ in the flow direction can be either upstream ($p_x^* > 0$) or downstream ($p_x^* < 0$), as determined by the function $\Omega_{x'}^{(ax)}(p_z, h/R)$; both cases are shown. When $p_z = 0$, all contributions to $\dot{\mathbf{p}}$ are in the $\hat{z}$ direction (see Eqs. 2-5 and the associated discussion in the main text). At the fixed point, all contributions to $\dot{\mathbf{p}}$ balance, as shown by the arrows, so that $\dot{\mathbf{p}} = 0$. Note that this fixed point always occurs in pairs related by mirror symmetry across the shear plane; we show the state with $p_y^* > 0$.

We supply some additional details concerning the fixed point solutions to Eqs. 2-5 of the main text. In a *rheotactic* state, the particle orientation vector is within the plane of shear. The component of the orientation vector in the flow direction is either upstream (as shown) or downstream. Clearly, this state, although interesting, is not relevant to the experiments considered here (see Ref. 21 for a detailed analysis of it.) In a *log rolling* state, the particle orientation $\mathbf{p}$ is aligned in the vorticity direction $\pm\hat{y}$ (we show $\mathbf{p} = -\hat{y}$). This state nearly matches the experimental observations, but – crucially – does not account for the slight downstream orientation noted above. In *planar alignment*, the particle orientation is within the plane of the wall, so that $p_z^* = 0$. The component in the flow direction can be either be

upstream ($p_x < 0$) or downstream ($p_x > 0$); to emphasize this fact, we show both cases. Additionally, for both cases, we show the particle as having $p_y^* > 0$. Due to the mirror symmetry of Eqs. 2-5 of the main text with respect to $p_y = 0$, a the planar alignment fixed point always occurs in pairs $(p_x, \pm p_y^*, 0, h^*)$.

The planar alignment fixed point shows excellent qualitative agreement with the experiment observations: planar alignment occurs as two states that are related by mirror symmetry across the shear plane, and the orientation of a particle in planar alignment has components in both the flow and vorticity directions. Owing to the non-zero vorticity component, the particle migrates across streamlines. We therefore consider planar alignment in mathematical detail. As discussed in Ref. 21, where this fixed point was designated "case (ii)", it has

$$p_x^* = \frac{2\Omega_{x'}^{(ax)}(p_z^* = 0, h^*)}{\dot{\gamma} f(h^*/R)} \tag{14}$$

and $p_z^* = 0$. Note that whether the particle orientation is upstream or downstream depends on the sign of $\Omega_{x'}^{(ax)}(p_z^* = 0, h^*)$. $p_y^*$ follows from $|\mathbf{p}| = 1$, so that $p_y^* = \pm\sqrt{1 - p_x^{*2}}$. The height $h^*$ is implicitly determined by the condition $U_{z'}^{(ax)}(p_z, h) = 0$. If this condition cannot be satisfied, the fixed point does not exist. Likewise, $|\mathbf{p}| = 1$ gives an additional condition for the existence of the fixed point, $\left|\frac{2\Omega_{x'}^{(ax)}(p_z^* = 0, h^*)}{\dot{\gamma} f(h^*/R)}\right| \leq 1$.

## S10. LINEAR STABILITY OF PLANAR ALIGNMENT

As discussed in the main text, we consider small perturbations $\tilde{\mathbf{f}}$ away from the planar alignment fixed point $\mathbf{f}^*$, such that $\mathbf{f} = \mathbf{f}^* + \tilde{\mathbf{f}}$. We obtain linearized governing equations $\dot{\tilde{\mathbf{f}}} = \mathbf{J}\tilde{\mathbf{f}}$, where $J_{ij} = \left.\frac{\partial \dot{f}_i}{\partial f_j}\right|_{p_z^* = 0, H}$ is the Jacobian matrix. We calculate:

$$\mathbf{J} = \begin{pmatrix} 0 & 0 & \frac{1}{2}\dot{\gamma} f(h^*/R) - \Omega_{x'}^{(ax)}(p_z^*, h^*/R)p_x^* & 0 \\ 0 & 0 & -\Omega_{x'}^{(ax)}(p_z^*, h^*/R)p_y^* & 0 \\ -\frac{1}{2}\dot{\gamma} f(h^*/R) & 0 & \left.\frac{\partial \Omega_{x'}^{(ax)}}{\partial p_z}\right|_{p_z^*=0, h=h^*} & -\frac{1}{2}\dot{\gamma} p_x^* \left.\frac{\partial f}{\partial h}\right|_{h=h^*} + \left.\frac{\partial \Omega_{x'}^{(ax)}}{\partial h}\right|_{p_z^*=0, h=h^*} \\ 0 & 0 & \left.\frac{\partial U_{z'}^{(ax)}}{\partial p_z}\right|_{p_z^*=0, h=h^*} & \left.\frac{\partial U_{z'}^{(ax)}}{\partial h}\right|_{p_z^*=0, h=h^*} \end{pmatrix},$$

which is compactly written as

$$\dot{\tilde{p}}_x = J_{13}\tilde{p}_z \tag{15}$$
$$\dot{\tilde{p}}_y = J_{23}\tilde{p}_z \tag{16}$$
$$\dot{\tilde{p}}_z = J_{31}\tilde{p}_x + J_{33}\tilde{p}_z + J_{34}\tilde{h} \tag{17}$$
$$\dot{\tilde{h}} = J_{43}\tilde{p}_z + J_{44}\tilde{h}. \tag{18}$$

We assume constant height $h = H$, so that $\dot{h} = 0$. We take the first derivative of Eq. 17, and substitute Eq. 16 to obtain

$$\ddot{\tilde{p}}_z = J_{31}J_{13}\tilde{p}_z + J_{33}\dot{\tilde{p}}_z \tag{19}$$

which is identical to the equation for a linearly damped harmonic oscillator. Substituting for $J_{31}$, $J_{23}$, and $J_{33}$, we obtain

$$\ddot{\tilde{p}}_z = -\frac{1}{2}\dot{\gamma} f(H/R)\left(\frac{1}{2}\dot{\gamma} f(H/R) - \Omega_{x'}^{(ax)}(p_z^*, H/R)p_x^*\right)\tilde{p}_z + \left.\frac{\partial \Omega_{x'}^{(ax)}}{\partial p_z}\right|_{p_z^*=0, H}\dot{\tilde{p}}_z \tag{20}$$

$$\ddot{\tilde{p}}_z = -\left(\frac{1}{2}\dot{\gamma} f(H/R)\right)^2\left(1 - \frac{2\Omega_{x'}^{(ax)}(p_z^*, H/R)p_x^*}{\dot{\gamma} f(H/R)}\right)\tilde{p}_z + \left.\frac{\partial \Omega_{x'}^{(ax)}}{\partial p_z}\right|_{p_z^*=0, H}\dot{\tilde{p}}_z \tag{21}$$

Substituting Eq. 14,

$$\ddot{\tilde{p}}_z = -\left(\frac{1}{2}\dot{\gamma}f(H/R)\right)^2 \left(1-p_x^{*2}\right)\tilde{p}_z + \frac{\partial \Omega_{x'}^{(ax)}}{\partial p_z}\bigg|_{p_z^*=0,H}\dot{\tilde{p}}_z \tag{22}$$

$$\ddot{\tilde{p}}_z = -\left(\frac{1}{2}\dot{\gamma}f(H/R)p_y^*\right)^2 \tilde{p}_z + \frac{\partial \Omega_{x'}^{(ax)}}{\partial p_z}\bigg|_{p_z^*=0,H}\dot{\tilde{p}}_z. \tag{23}$$

For an inactive, bottom-heavy particle driven by flow, we have

$$\frac{\partial \Omega_{x'}^{(ax)}}{\partial p_z}\bigg|_{p_z^*=0,H} = 0, \tag{24}$$

and we can easily construct an analytical solution to Eq. 23: $p_z(t) = A\cos(\omega_0 t - \psi)$, with $\psi = \operatorname{atan}(\dot{p}_z(0)/\omega_0 p_z(0))$, $A = \left(p_z^2(0) + \frac{\dot{p}_z^2(0)}{\omega_0^2}\right)^{1/2}$, and $\omega_0 = \frac{1}{2}\dot{\gamma}f(H/R)p_y^*$. We evaluate the parameters $A$, $\omega_0$ and $\psi$ numerically, using BEM calculations. For the trajectory in Fig. 5a and Fig. 6a of the main text, we plot the analytical solution as a red dashed line in Fig. 6e of the main text. The full numerical result and the solution to the linearized equations show good agreement, with some deviation due to the large amplitude of the oscillation.

For an active particle, the analytical solution to Eq. 23 equation is damped: $p_z(t) = Ae^{-\nu t/2}\cos(\omega_1 t - \psi)$, where $\nu = -\frac{\partial \Omega_{x'}^{(ax)}}{\partial p_z}\bigg|_{p_z=0,h=H}$, $\omega_1 = (\omega_0 - \nu^2/4)^{1/2}$, $A = \left[p_z^2(0) + \left(\frac{\dot{p}_z(0)+\nu p_z(0)/2}{\omega_1}\right)^2\right]^{1/2}$, and $\psi = \operatorname{atan}\left(\frac{\dot{p}_z(0)+\nu p_z(0)/2}{\omega_1 p_z(0)}\right)$.

Again, we evaluate the parameters $\nu$, $A$, $\psi$, and $\omega_1$ numerically using the BEM. For the trajectory in Fig. 5b and Fig. 6b of the main text, we plot the analytical solution in Fig. 6f of the main text as a dashed red line for when $p_z$ has low amplitude ($t/T_0 > 200$), where we expect the linearized equations to be valid. We find good agreement with the numerics in this regime.

## S11. STEADY ANGLE OF A PARTICLE AS A FUNCTION OF FLOW RATE

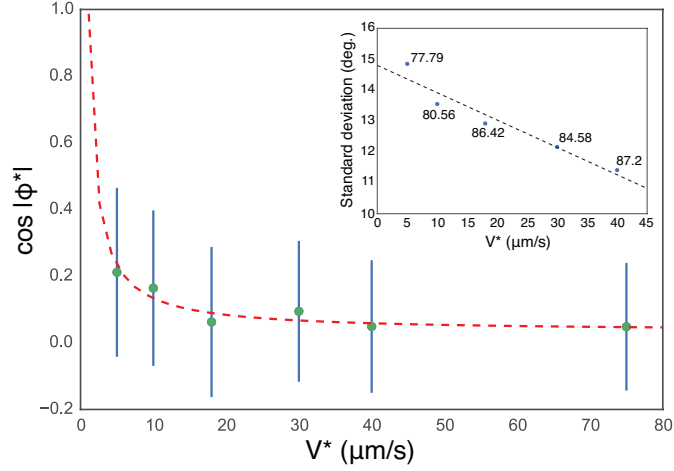

FIG. S11. The mean $p_x = |\cos(\phi*)|$ plotted as a function of $V^*$. Data was obtained from experiments with $R = 2.5$ $\mu m$ particles at different flow rates. The dotted line is a fit to the predicted scaling relation. We clearly recover the asymptotic prediction of the scaling relation, i.e., that $p_x = |\cos(\phi*)|$ approaches zero for fast flow. (Inset) The dependence of standard deviation of $\phi^*$ as a function of $V^*$ is approximately linear. Numbers show the mean $\phi^*$.

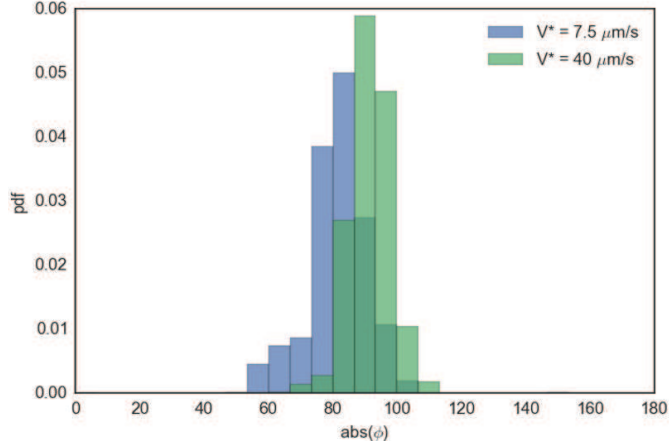

FIG. S12. Probability distribution of $|\phi|$ plotted for two different flow velocities (corresponding to two data points in Fig. S11) shows a clear shift of the peak position $\phi^*$ towards 90° at higher flow rates ($V_p \approx 6$ $\mu m/s$, $R = 2.5$ $\mu m$).

## S12. CALCULATION OF GRAVITATIONAL CONTRIBUTION TO PARTICLE MOTION

We use the "eggshell" model of Campbell and Ebbens for the shape of the cap, taking the cap thickness to vary smoothly from zero at the particle "equator" to a maximum thickness of $t$ at the active pole [4]. For this geometry, the buoyancy force on the particle is $\mathbf{F}_g = \frac{2}{3}\pi t R^2 (\rho_{cap} - \rho_{fluid}) g \hat{z}' + \frac{4}{3}\pi R^3 (\rho_{core} - \rho_{fluid}) g \hat{z}'$, where $\rho_{fluid}$ is the mass density of the fluid, $\rho_{cap}$ is the mass density of the catalytic cap, and $\rho_{core}$ is the mass density of the inert core. The torque from bottom-heaviness is $\boldsymbol{\tau}_g = \frac{2}{3}\pi R^2 t \rho_{cap} g y_{cm} \hat{x}'$, where $y_{cm} \approx \frac{3}{4}(R+t)$. The boundary conditions for the fluid velocity are $\mathbf{u} = 0$ on the wall and $\mathbf{u} = \mathbf{U}^{(g)} + \boldsymbol{\Omega}^{(g)} \times (\mathbf{x}' - \mathbf{x}'_p)$ on the particle surface. Through a standard calculation [11], the generalised velocity $\mathbf{V}^{(g)} \equiv (\mathbf{U}^{(g)}, \boldsymbol{\Omega}^{(g)})$ is obtained as $\mathbf{V}^{(g)} = \mathbf{R}^{-1}\mathfrak{F}^{(g)}$, where $\mathfrak{F}^{(g)} \equiv (\mathbf{F}^{(g)}, \tau^{(g)})$ and $\mathbf{R}$ is the hydrodynamic resistance tensor. Since we take the particle to have a constant height $h \sim H$, we consider

the gravitational force on the particle to be balanced by an electrostatic force from the wall. Hence, in calculating particle trajectories, we use only $\mathbf{\Omega}^{(g)}$. Regarding material parameters, we take the particle to have a silica core with radius of $R = 2.5 \ \mu m$ and density $\rho_{core} = 2648 \ kg/m^3$, and platinum cap with maximum thickness $t = 10 \ nm$ and density $\rho_{cap} = 21450 \ kg/m^3$. The suspending fluid is mostly water, with $\rho_{fluid} \approx 1000 \ kg/m^3$.

## S13. MOTION OF PARTICLE IN THREE DIMENSIONS

With a brief example, we show that the particle can attain a "'planar alignment'" steady state, i.e., exhibit cross-stream migration, even if the height $h$ is allowed to change according to Eq. 5 in the main text. In Fig. S13, we show a trajectory for a particle with initial height $h/R = 5$ and orientation $\phi = 120°$ and $\theta = 100°$. The particle quickly attains a steady height $h/R = 1.23$ and orientation $\theta = 90°$ and $\phi = 83°$.

Here, we have included a short-ranged repulsive force between the particle and the wall to prevent the particle from "crashing" into the wall, i.e., moving below $h/R = 1$. Motivated by electrostatics, we use a force of the form:

$$F^{es}(h/R) = \frac{F_0 \, e^{-kR(h/R-1)/2}}{\cosh(kR(h/R-1)/2)} \tag{25}$$

Here, we take $kR \sim 200$ and $F_0/6\pi\mu R U_0 \sim 400$. The length scale of the force $k^{-1} \sim 0.005R$. The short-ranged character of the force is illustrated by the fact that $U_z^{es}$, its contribution to the vertical velocity of the particle, is less than $\sim 0.02 U_0$ for $h/R \geq 1.04$. The repulsive force has therefore negligible role in determination of the steady height and orientation of the particle.

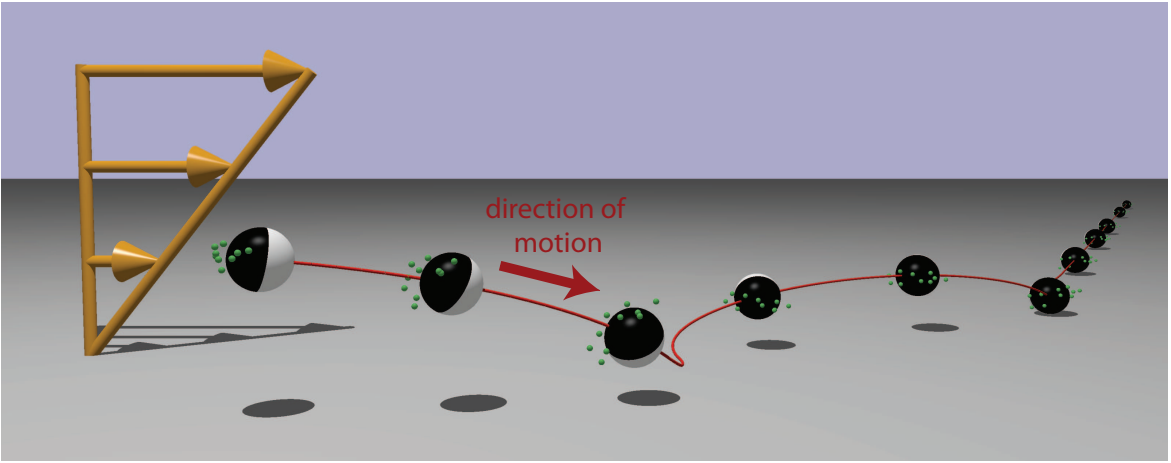

FIG. S13. Three-dimensional trajectory of a bottom-heavy, chemically active particle in a shear flow, with parameters as described in the "Methods" section. Additionally, we include a short-ranged repulsive interaction between the wall and the particle to prevent the particle from "crashing" into the wall. From an initial height $h/R = 5$ and orientation $\phi = 120°$ and $\theta = 100°$, the particle quickly attains a steady height $h/R = 1.23$ and orientation $\theta = 90°$ and $\phi = 83°$.

## S14. BROWNIAN DYNAMICS SIMULATIONS

The simulation procedure is as follows. Before the simulation, we compute the (height dependent) non-dimensionalized components of the diffusion tensor: $\tilde{D}_{||}^t \equiv D_{||}^t/D_0^t$, $\tilde{D}_{\perp}^t \equiv D_{\perp}^t/D_0^t$, $\tilde{D}_{||}^r \equiv D_{||}^r/D_0^r$, $\tilde{D}_{\perp}^r \equiv D_{\perp}^r/D_0^r$, and $\tilde{D}^{tr} \equiv D_{tr}^r/RD_0^r$. Here, the symbols "$||$" and "$\perp$" denote directions parallel to and normal to the wall, respectively; "$r$" indicates a rotational component; "$t$" indicates a translational component; and "$tr$" indicates the component coupling transla- tion and rotation. Here, $D_0^r = k_bT/8\pi\eta R^3$ is the free space rotational diffusion coefficient. We then compute the components $B_{ij}$ defined by the Cholesky decomposition $\tilde{D}_{ij} = B_{ik}B_{jk}$ using the analytical expressions in Ref. 55. We then run the simulation. At each time step, we define $\mathbf{p}_0 = \mathbf{p}(t)$. We first compute the deterministic change in the orientation vector $\dot{\mathbf{p}}$ from Eqs. 2-5 of the main text. We then generate a vector $\mathbf{b}$ of six Gaussian distributed independent random variables with $\langle b_i \rangle = 0$ and $\langle b_i b_j \rangle = 2Pe_p^{-1}\Delta\tilde{t}\delta_{ij}$, where $\Delta\tilde{t} \equiv \Delta t/T_0$ is the size of the time step,

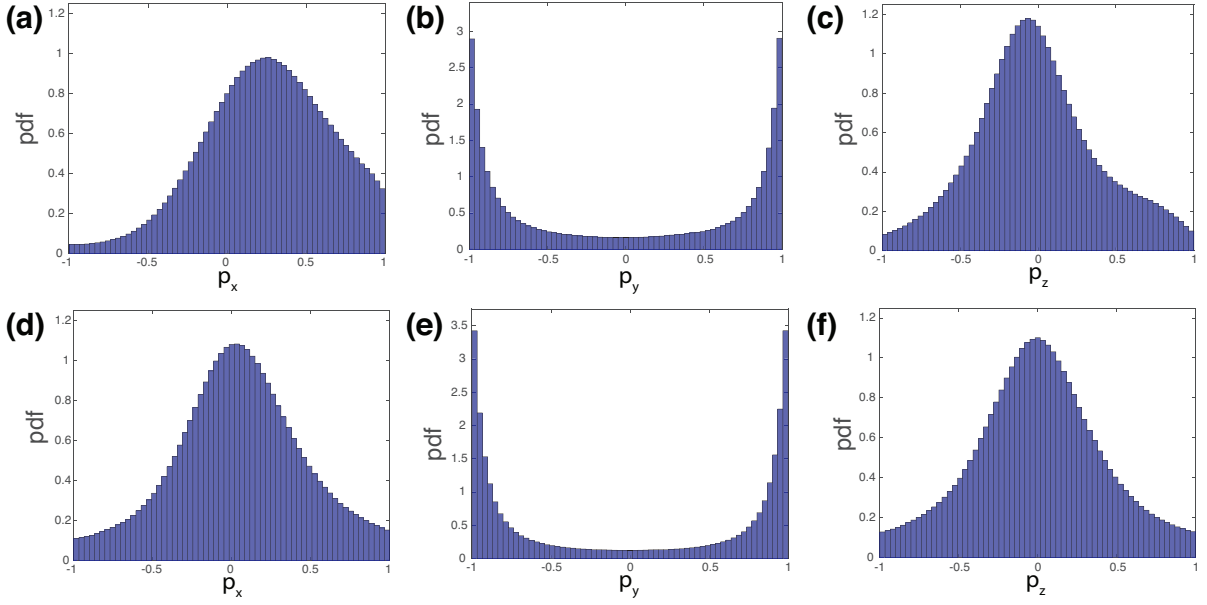

FIG. S14. Probability distribution functions for components of the particle orientation vector **p**, obtained from stochastic numerical simulations as described in the main text. In panels (a) through (c), the dimensionless shear rate is $\dot{\gamma}R/U_0 = 0.1$; in panels (d) through (f), it is $\dot{\gamma}R/U_0 = 0.5$.

and $i = 1, ..., 6$. From **b**, we generate a vector $\mathbf{s} = \mathbf{B} \cdot \mathbf{b}$. The last three components of **s** ($i = 4, 5, 6$) define the vector $\mathbf{s}^r$. We then compute

$$\mathbf{p}' = \mathbf{p}_0 + \dot{\mathbf{p}}\Delta \tilde{t} - \frac{3}{4}(\tilde{D}^r_\perp - \tilde{D}^r_{||})(p_{z,0}^2 \mathbf{p}_0 - p_{z,0}\hat{z})Pe_p^{-1}\Delta \tilde{t} + \mathbf{s}^r \times \mathbf{p}_0. \quad (26)$$

Finally, we normalize the length of $\mathbf{p}'$ to obtain the orientation vector at the next time step: $\mathbf{p}(t + \Delta t) = \mathbf{p}'/|\mathbf{p}'|$. The distributions shown in Fig. 5 of the main text are computed for an ensemble of fifty trajectories with random initial orientations and using a timestep of $\Delta \tilde{t} = 0.005$, with the run time of each trajectory $T_{run}/T_0 = 50000$. All trajectories are sampled after an initial transient period of $T_{start}/T_0 = 2500$, allowing the particle orientations to evolve from the randomly distributed initial orientation; this corresponds to $(T_{start}/T_0) Pe_p^{-1} = 5$ characteristic diffusion times.

## S15. LEGENDS FOR MOVIES S1 TO S3

### 1. Movie S1: Silica-Pt active Janus particles in the absence of any external shear flow

Upon addition of hydrogen peroxide the silica-Pt colloids self-propel along the surface of the channel. The $\theta$ is fixed at $\theta \approx 90°$ and the particles remain in-plane. The direction of self-propulsion is away from the Pt cap (seen as the dark hemisphere). Real time movie.

### 2. Movie S2: Silica-Pt Janus inactive particles in a shear flow (no hydrogen peroxide).

The silica-Pt Janus colloids are in a suspension of water (no activity). Initially, they all appear as black circles as they are sedimented with their Pt caps down. This orientation is induced by the bottom heaviness of the particles. Once the flow is started, the particles are advected by the shear flow, tumbling, while they translate in the flow direction. Intermediate states during the tumbling are visible due to the contrast between the Pt and the silica halves. Real time movie.

3. *Movie S3: Silica-Pt Janus active particles in a shear flow.*

A flow of hydrogen peroxide is introduced in a capillary with the silica-Pt Janus particles. The particles have a stable orientation with respect to the bottom surface at $\theta \approx 90°$. Additionally the particles also achieve a stable orientation with respect to the flow direction at $\phi \approx +90°$ or $-90°$. Since the particles self-propel away from their Pt caps, they migrate across streamlines. Real time movie.



\end{document}